\documentstyle[12pt]{article}

\advance \topmargin by -\headheight
\advance \topmargin by -\headsep
     
\evensidemargin \oddsidemargin
\begin{document}

\newcommand\twomat[4]{\left(\begin{array}{cc}  
{#1} & {#2} \\ {#3} & {#4} \end{array} \right)}
\newcommand\twocol[2]{\left(\begin{array}{cc}  
{#1} \\ {#2} \end{array} \right)}

\newcommand{\sect}[1]{\setcounter{equation}{0}\section{#1}}
\renewcommand{\theequation}{\thesection.\arabic{equation}}

\topmargin -.6in
\def\nonu{\nonumber}
\def\rf#1{(\ref{eq:#1})}
\def\lab#1{\label{eq:#1}} 
\def\br{\begin{eqnarray}}
\def\er{\end{eqnarray}}
\def\be{\begin{equation}}
\def\ee{\end{equation}}
\def\0{\nonumber}
\def\lb{\lbrack}
\def\rb{\rbrack}
\def\({\left(}
\def\){\right)}
\def\v{\vert}
\def\bv{\bigm\vert}
\def\lskip{\vskip\baselineskip\vskip-\parskip\noindent}
\relax
\newcommand{\nit}{\noindent}
\newcommand{\ct}[1]{\cite{#1}}
\newcommand{\bi}[1]{\bibitem{#1}}
\def\a{\alpha}
\def\b{\beta}
\def\ca{{\cal A}}
\def\cm{{\cal M}}
\def\cn{{\cal N}}
\def\cf{{\cal F}}
\def\d{\delta} 
\def\D{\Delta}
\def\eps{\epsilon}
\def\g{\gamma}
\def\G{\Gamma}
\def\grad{\nabla}
\def\h{ {1\over 2}  }
\def\hc{\hat{c}}
\def\hd{\hat{d}}
\def\hg{\hat{g}}
\def\hp{ {+{1\over 2}}  }
\def\hm{ {-{1\over 2}}  }
\def\k{\kappa}
\def\l{\lambda}
\def\L{\Lambda}
\def\lg{\langle}
\def\m{\mu}
\def\n{\nu}
\def\o{\over}
\def\om{\omega}
\def\O{\Omega}
\def\p{\phi}
\def\pa{\partial}
\def\pr{\prime}
\def\ra{\rightarrow}
\def\rh{\rho}
\def\rg{\rangle}
\def\s{\sigma}
\def\t{\tau}
\def\th{\theta}
\def\ti{\tilde}
\def\wti{\widetilde}
\def\inte{\int dx }
\def\xb{\bar{x}}
\def\yb{\bar{y}}

\def\tr{\mathop{\rm tr}}
\def\Tr{\mathop{\rm Tr}}
\def\partder#1#2{{\partial #1\over\partial #2}}
\def\ds{{\cal D}_s}
\def\wtwo{{\wti W}_2}
\def\lie{{\cal G}}
\def\alie{{\widehat \lie}}
\def\dlie{{\cal G}^{\ast}}
\def\elie{{\widetilde \lie}}
\def\edlie{{\elie}^{\ast}}
\def\hlie{{\cal H}}
\def\wlie{{\widetilde \lie}}

\def\rlx{\relax\leavevmode}
\def\inbar{\vrule height1.5ex width.4pt depth0pt}
\def\IZ{\rlx\hbox{\sf Z\kern-.4em Z}}
\def\IR{\rlx\hbox{\rm I\kern-.18em R}}
\def\IC{\rlx\hbox{\,$\inbar\kern-.3em{\rm C}$}}
\def\one{\hbox{{1}\kern-.25em\hbox{l}}}

\def\PRL#1#2#3{{\sl Phys. Rev. Lett.} {\bf#1} (#2) #3}
\def\NPB#1#2#3{{\sl Nucl. Phys.} {\bf B#1} (#2) #3}
\def\NPBFS#1#2#3#4{{\sl Nucl. Phys.} {\bf B#2} [FS#1] (#3) #4}
\def\CMP#1#2#3{{\sl Commun. Math. Phys.} {\bf #1} (#2) #3}
\def\PRD#1#2#3{{\sl Phys. Rev.} {\bf D#1} (#2) #3}
\def\PLA#1#2#3{{\sl Phys. Lett.} {\bf #1A} (#2) #3}
\def\PLB#1#2#3{{\sl Phys. Lett.} {\bf #1B} (#2) #3}
\def\JMP#1#2#3{{\sl J. Math. Phys.} {\bf #1} (#2) #3}
\def\PTP#1#2#3{{\sl Prog. Theor. Phys.} {\bf #1} (#2) #3}
\def\SPTP#1#2#3{{\sl Suppl. Prog. Theor. Phys.} {\bf #1} (#2) #3}
\def\AoP#1#2#3{{\sl Ann. of Phys.} {\bf #1} (#2) #3}
\def\PNAS#1#2#3{{\sl Proc. Natl. Acad. Sci. USA} {\bf #1} (#2) #3}
\def\RMP#1#2#3{{\sl Rev. Mod. Phys.} {\bf #1} (#2) #3}
\def\PR#1#2#3{{\sl Phys. Reports} {\bf #1} (#2) #3}
\def\AoM#1#2#3{{\sl Ann. of Math.} {\bf #1} (#2) #3}
\def\UMN#1#2#3{{\sl Usp. Mat. Nauk} {\bf #1} (#2) #3}
\def\FAP#1#2#3{{\sl Funkt. Anal. Prilozheniya} {\bf #1} (#2) #3}
\def\FAaIA#1#2#3{{\sl Functional Analysis and Its Application} {\bf #1} (#2)
#3}
\def\BAMS#1#2#3{{\sl Bull. Am. Math. Soc.} {\bf #1} (#2) #3}
\def\TAMS#1#2#3{{\sl Trans. Am. Math. Soc.} {\bf #1} (#2) #3}
\def\InvM#1#2#3{{\sl Invent. Math.} {\bf #1} (#2) #3}
\def\LMP#1#2#3{{\sl Letters in Math. Phys.} {\bf #1} (#2) #3}
\def\IJMPA#1#2#3{{\sl Int. J. Mod. Phys.} {\bf A#1} (#2) #3}
\def\AdM#1#2#3{{\sl Advances in Math.} {\bf #1} (#2) #3}
\def\RMaP#1#2#3{{\sl Reports on Math. Phys.} {\bf #1} (#2) #3}
\def\IJM#1#2#3{{\sl Ill. J. Math.} {\bf #1} (#2) #3}
\def\APP#1#2#3{{\sl Acta Phys. Polon.} {\bf #1} (#2) #3}
\def\TMP#1#2#3{{\sl Theor. Mat. Phys.} {\bf #1} (#2) #3}
\def\JPA#1#2#3{{\sl J. Physics} {\bf A#1} (#2) #3}
\def\JSM#1#2#3{{\sl J. Soviet Math.} {\bf #1} (#2) #3}
\def\MPLA#1#2#3{{\sl Mod. Phys. Lett.} {\bf A#1} (#2) #3}
\def\JETP#1#2#3{{\sl Sov. Phys. JETP} {\bf #1} (#2) #3}
\def\JETPL#1#2#3{{\sl  Sov. Phys. JETP Lett.} {\bf #1} (#2) #3}
\def\PHSA#1#2#3{{\sl Physica} {\bf A#1} (#2) #3}
\def\PHSD#1#2#3{{\sl Physica} {\bf D#1} (#2) #3}

\begin{titlepage}
\vspace*{-2 cm}
\noindent
\begin{flushright}
\end{flushright}

\vskip 1 cm
\begin{center}
{\Large\bf Electrically Charged Topological Solitons  } \vglue 1  true cm
{ J.F. Gomes}, E. P. Gueuvoghlanian,
 { G.M. Sotkov} and { A.H. Zimerman}\\

\vspace{1 cm}

{\footnotesize Instituto de F\'\i sica Te\'orica - IFT/UNESP\\
Rua Pamplona 145\\
01405-900, S\~ao Paulo - SP, Brazil}\\
jfg@ift.unesp.br, gueuvogh@ift.unesp.br, sotkov@ift.unesp.br, zimerman@ift.unesp.br\\

\vspace{1 cm}

\end{center}

\normalsize
\vskip 0.2cm

\begin{center}
{\large {\bf ABSTRACT}}\\
\end{center}
\noindent
Two new families of T-Dual integrable models of dyonic type are constructed. They represent
specific $A_n^{(1)}$ singular Non-Abelian Affine Toda models having $U(1)$ global symmetry.  
Their 1-soliton spectrum contains both neutral   and $U(1)$
charged  topological solitons  sharing the main properties of 4-dimensional
Yang-Mills-Higgs monopoles and dyons.  The semiclassical quantization of these solutions as well
as  the exact counterterms and the coupling constant renormalization are studied.
\noindent

\vglue 1 true cm

\end{titlepage}

\sect{Introduction}

In search for new nonperturbative methods of quantization of four dimensional 
(nonsupersymmetric) $SU(n+1)$ QCD (and its bosonic part- the
Yang-Mills-Higgs (YMH) model ), the relationship between 4-D selfdual Yang-Mills (SDYM) theories and certain 2-D 
integrable models (IM's) \cite{ward}
deserves special attention.  As it is well known the dimensional (and symmetries)
 reduction of 4-D SDYM to lower dimensional (D=1,2,3)
IM's (see \cite{ablo} for a review) provides effective methods for the construction 
of a large class of exact (classical ) solutions of the 4-D gauge theory,
including spherical symmetric monopoles \cite{leznov}, \cite{ganoulis}, instantons \cite{witt5}, domain walls (DW) \cite{shiff},
\cite{gau} etc.  Among them the solitons of
relativistic 2-D IM's are of particular interest:
\begin{itemize}
\item They are expected to describe  (at least in large n limit ) DW's solutions of 4-D $SU(n+1)$ YMH theory;
\item Their exact quantization is known to be related to the centerless affine quantum group $U_q(A_n^{(1)})$ (or $U_q(A_{2n}^{(2)}), 
U_q(A_{2n-1}^{(2)})$ ) \cite{bernard}.
\end{itemize} 
Whether $U_q(A_n^{(1)})$ (or its dual ) takes place in the description of the
 quantum DW's and the other ``strong coupling'' solutions of
$SU(n+1)$ YMH theory - monopoles, dyons and strings - is an open problem.  
There exists however a hint in favour of such conjectute,
namely:
The space of classical solutions of 4-D $SU(n+1)$ SDYM (and BPS solutions of YMH) has
 the affine  $\hat {SU}(n+1)$ loop algebra as dynamic
symmetry \cite{dolan2}, which coincides with the classical limit $q \rightarrow 1 (h \rightarrow 0)$ of the
 quantum soliton symmetry group
$U_q(A_n^{(1)})$ \cite{bernard}.


 The domain walls that appear in the string (i.e. D-branes ) description of 4-D gauge theories \cite{shiff}, \cite{gau} 
 manifest properties a bit more
 involved than the simplest topological solitons of say, sine-Gordon or abelian affine Toda models. 
  For instance, they may carry certain $U(1)$ charges
 or require nonmaximal breaking of $SU(n+1)$ gauge group, say $SU(k)\otimes U(1)^{n-k+1}$ ( 
 i.e. $k$-coincident D-branes) \cite{gau}.  The
 question arises whether one can find 2-D IM's whose soliton solutions possess all DW's characteristics.
 The
problem we address in the present paper is to construct  such {\it dyonic } integrable models
\footnote{The Sine Gordon (SG) and the $\hat G_n$-abelian affine Toda model are examples of {\it
magnetic } type IM having {\it neutral } topological solitons.  The Complex Sine Gordon ( or
Lund-Regge) model \cite{lund1},\cite{dorey},\cite{devega} and its homogeneous space SG
 generalizations \cite{pousa} represent the
{\it electric} type IM's, whose solitons are $U(1)$ (or $U(1)^r$ ){\it charged} but {\it
nontopological}.}
  admiting $U(1)$-{\it charged topological solitons
} and CPT-violating term.  We shall study the {\it solitonic spectrum}  (classical and quantum )
of the following {\it new family } of (CPT-noninvariant) $A_n^{(1)}$ dyonic IM's ( $n >1$ ) 
\be
{\cal L}_a = {1\o 2}\sum_{i,k=1}^{n-1}\eta_{ik}\pa \varphi_i \bar \pa \varphi_k + 
{{\pa \chi \bar \pa
\psi e^{-\b \varphi_1}}\o {\Delta}} -  V_a 
\label{1.1}
\ee
\br
V_a = {m^2 \o \b^2} \( \sum_{i=1}^{n-1} e^{\b (\varphi_{i-1} + \varphi_{i+1} 
- 2 \varphi_i )} +
e^{\b (\varphi_{1} + \varphi_{n-1})}(1 + \b^2 \psi \chi e^{-\b \varphi_1})- n\)
\nonumber
\er
where $\eta_{ik}= 2 \d_{i,k} - \d_{i, k-1} - \d_{i, k+1}, \;\; i,k = 1, 
\cdots n-1$, $\Delta = 1+
\b^2 {{n+1}\o {2n}} \psi \chi e^{-\b \varphi_1}$, $\varphi_0 = \varphi_n = 0$, 
$\pa = \pa_t + \pa _x,
\bar \pa = \pa_t - \pa_x $ and $\b^2 = -{{2 \pi}\o K}$.The IM's  (\ref{1.1})
are invariant
 under global $U(1)$
transformation, 
\br
\psi^{\pr} = e^{-ia\b^2} \psi, \quad \chi^{\pr} = e^{ia\b^2} \chi, \quad \varphi_i^{\pr} = \varphi_i
\nonumber
\er
where $a$ is a real constant.They  represent a specific  mixture of the Lund-Regge model interacting
with the $A_n^{(1)}$-abelian Toda theory.  It is worthwhile to mention that
 the above dyonic
models appear to be an  appropriate integrable deformation of the recently constructed
$V_{n+1}^{(1)}$-algebra minimal models \cite{annals} (i.e. the singular $A_n$-Non Abelian (NA) Toda
theories):
\be
{\cal L} = {\cal L}_{conf} - {{m^2}\o {\b^2}} V_{def}, \quad V_{def} =e^{\b (\varphi_1 +
\varphi_{n-1})}(1 + \b^2 \psi \chi e^{-\b \varphi_1})
\label{1.2}
\ee
The CPT-violating term $e^{-\b  \varphi_1} \Delta^{-1} \eps^{\mu \nu} \pa_{\mu} \psi \pa_{\nu}\chi $
is  hidden in the second term of (\ref{1.1})
\br
{{e^{-\b \varphi_1}}\o { \Delta}} \bar  \pa \psi \pa \chi = {{e^{-\b \varphi_1}}\o { \Delta}}
\(g^{\mu \nu} \pa_{\mu} \psi \pa_{\nu}\chi + \eps^{\mu \nu} \pa_{\mu} \psi \pa_{\nu}\chi\)
\nonu 
\er
It has the same origin as in the conformal theory ${\cal L}_{conf}$, i.e.  the {\it axial}
 gauging of the $U(1)$ subgroup spanned by $\l_1 \cdot H$.  The CPT-invariant partner of
(\ref{1.1}) is the following  vector type $A_n^{(1)}$-dyonic model:
\br
{\cal L}_{v}&=&{1\o {\b^2}}\Biggl({1\o 2} \sum_{i=1}^{n-1} [ \pa ln c_i \bar \pa
 ln (c_i c_{i+1} \cdots c_{n-1})
+
\bar \pa ln c_i  \pa ln (c_i c_{i+1} \cdots c_{n-1})] \nonu \\
 &-&{1\o 2} {{\pa A \bar \pa B +\pa B \bar
\pa A}\o {1- AB}} - \b^2 V_{v}\Biggr) 
\label{1.3}
\er
\br
V_v = {{m^2}\o \b^2}\( A c_1^2c_2 \cdots c_{n-1} + {{B}\o {c_1 c_2 
\cdots c_{n-2}c_{n-1}^2}} +
{{c_2}\o c_{1}} + {{c_3}\o c_{2}} + \cdots +{{c_{n-1}}\o c_{n-2}}  -n \)
\nonu 
\er
It is obtained by {\it vector } gauging of the same $U(1)$ subgroup and 
appears to be T-dual to the `` axial '' model (\ref{1.1}) \cite{ggsz1}.  The
ungauged
integrable model giving rise to both models (\ref{1.1}) and (\ref{1.3}) has the following
Lagrangean \footnote{Note that the $n=2$, (i.e. $SL(3,R)$) IM is nothing but
the (thermal operator ) {\it integrable perturbation } of the
Bershadsky-Polyakov $W_3^{(2)}$ minimal models \cite{poly}.}
\be
{\cal L}_{u} ={1\o 2} \eta_{ik}\pa \varphi_i \bar \pa \varphi_k + {{(n+1)}\o {2n}}\pa R
\bar \pa R + \pa {\tilde \chi} \bar \pa{\tilde  \psi} e^{\b ({{(n+1)R}\o n}-\varphi_1)} - 
 V_{u}
\label{1.4}\nonu 
\ee
\br
V_u = {{m^2}\o \b^2}\( \sum_{i=1}^{n-1} e^{\b (\varphi_{i-1}+\varphi_{i+1} - 2\varphi_i)} +
e^{\b (\varphi_1 + \varphi_{n-1})}(1 + \b^2 {\tilde \psi} \tilde \chi e^{\b ({{(n+1)R}\o n}-\varphi_1)})-n\)
\nonu
\er
The corresponding action $S_u$  is invariant under chiral $U(1)$ gauge transformations: 
\br
 R^{\pr} = R+w(z) + \bar w (\bar z),\quad 
\tilde \psi^{\pr} = e^{-{{\b (n+1)w(z)}\o {n}}}\tilde \psi, \quad  \tilde \chi^{\pr} =
 e^{-{{\b (n+1)\bar w(\bar z)}\o {n}}}\tilde \chi
 \nonu
 \er
  where $\psi = \tilde \psi e^{{{(n+1)}\o {2n}}R},\chi = \tilde \chi e^{{{(n+1)}\o {2n}}R}$.

The {\it zero curvature} representation  and the proof of the {\it classical }
integrability of all  models (\ref{1.1}), (\ref{1.3}) and (\ref{1.4}), is
derived in Sect. 2.  Special attention is devoted to the discussion of  {\it
discrete symmetries } of (\ref{1.1})  and  (\ref{1.3}).  They are
crucial in the definition of the vacuum lattice, the spectra of the topological
charges $Q_k^{top.}$ and $Q_{\theta}$ as well as in the derivation of the first
order 1-soliton equations in Sect.3.  It is important to note that the first order system 
(\ref{3.61a})    has to be completed with a chain of {\it
algebraic relations }(\ref{3.8}) in order to determine a consistent vacua
Backlund transformation.  They represent 2-D analog of the YMH-model Bogomolny
equations and 
constitute our main tool in constructing 1-soliton solutions of the dyonic models
(\ref{1.1})  and  (\ref{1.3}) in Sects. 4 and 5.  Compared to the other
methods of construction of soliton solutions as ``vertex operators '' (or $\tau
$-functions, etc )\cite{olive1},\cite{babelon},\cite{luiz}  they have the advantage 
in providing a simple proof
of the {\it topological character } of the soliton's energy, momenta and action
$S_{1-sol}(A_n^{(1)})$.  It is shown in Sect. 3 that all conserved quantities
($ E^{ax}, P^{ax}, M^{ax}, Q_{el}^{ax}, Q_{\theta}^{ax}, Q_k^{top}, s^{ax}$) characterizing $U(1)$ charged
topological 1-solitons depend only on the asymptotics of the fields at $x
\rightarrow \pm \infty $ .Therefore their values can be calculated {\it without } the
knowledge of the explicit form of the solutions.

The {\it main result } of this paper is the explicit construction of the charged
topological 
1-solitons of the {\it axial} and {\it vector} models 
 (\ref{1.1})  and  (\ref{1.3})  presented in Sect. 4 and 5.  Their 
 {\it semiclassical} spectrum  (derived in Sect. 3,6 and 7) is shown to have the
 following promissing {\it dyonic }\footnote { to be compared with the semiclassical dyonic spectrum of
 YMH model (see for example sect.VI of ref.\cite{zamora}).} form
\br
 Q_{el}^{ax} = \b_0^2  ( j_{el} + {{\nu }\o {2\pi}}j_{\varphi}), \quad
 Q_{\theta}^{ax} = {{2\pi \nu  \g_0 }\o {\nu - {{8\pi}\o {\b_0^2}}}} j_{\varphi},
  \quad Q_{mag} = {{4\pi }\o {\b_0^2}}j_{\varphi}, \quad 
j_{\varphi} = 0, \pm 1,  \cdots ,\pm (n-1)  
\nonumber
\er
\br
 M_{j_{el}, j_{\varphi}}^{ax} = {{4m n}\o{\b_0^2}}|\sin {{{Q_{el}^{ax}}- {\b_0^2}Q_{mag}}\o {4n}}|,\quad 
 s^{ax} = {{(n-1)}\o {2\b_0^2}}\( Q_{el}^{ax}+ {\b_0^2}Q_{mag}\),
 \;\; \;\; j_{el} = 0, \pm 1, \cdots \nonu \\ 
 \label{1.5}
\er
( $\nu$ and  $\g_0 $ are arbitrary real numbers ; $\b = i\b_0$ )
for the ``axial'' model (\ref{1.1}) and a similar form with $(Q_{el}^{ax} ,
Q_{\theta}^{ax})\longrightarrow  ( Q_{\tilde \theta}^{vec}, Q_{el}^{vec})$ 
interchanged for its T-dual ``vector'' model (\ref{1.3}).  It is worthwhile  
 to mention that $j_{\varphi}$ (and $\nu, \g_0$)-dependent shifts of $Q_{el}^{ax}$ and $Q_{\theta}^{ax}$ come from the topological
 $\theta$-terms (\ref{2.16}) and (\ref{varl}) added to ${\cal L}_a$ and ${\cal L}_v$ respectively. 
  Our proof that the ``axial'' and ``vector ''
 models 1-solitons have equal masses, energies and spins (see Sect. 5) represents an important step in testing the `` off-critical'' (i.e.
 nonconformal ) T-duality.  An interesting property of the mass spectrum (\ref{1.5}) is that its large n limit 
 \br
 M^{ax} (n \rightarrow \infty ) = {{m\o {\b_0^2}}}| Q_{el}^{ax} - \b_0^2 Q_{mag} |
 \nonu
 \er
 coincides with the BPS bound for the masses of particular ($Q_{el}^{(2)} = 0 = Q_{mag}^{(1)}$) dyons of 4-D $N=2$ super YM theory, i.e.
 $M^{ax} (n \rightarrow \infty ) \sim Z_- = Q_{el}^{(1)} - Q_{m}^{(2)}$, \cite{gauntlett}.
 
 Sect. 7 contains  {\it preliminary } results
concerning the {\it exact quantization } of the $A_n^{(1)}$ dyonic models in
consideration. 
Our starting point is the path integral formulation of the models
(\ref{1.1})  and  (\ref{1.3}) as gauged $\hat {H_-}\backslash A_n^{(1)} /\hat
{H_+}$ two loop WZW model (see refs. \cite{ggsz2} ).  We next derive the exact
effective action for the $A_n^{(1)}$ dyonic integrable models following the
method \cite{tsey} developed for the corresponding conformal $\s $-models (i.e. $m=0$
case).  It turns out that the relevant {\it counterterms} as well as the finite
{\it renormalization } of the coupling constants $\b^2_{renorm}$ are identical to
those of the conformal models (i.e. $V_{def}=0$ in eqn. (\ref{1.2})).
One can further speculate that eqns. (\ref{1.5}) with $\b_0^2 \rightarrow 
 \b_{0,renorm}^2 = {{2\pi}\o {k-n-1}}$ represent the {\it exact quantum } 1-soliton spectrum.  Few hints
 in favour of such {\it conjecture} are given in sect. 7.3.  The {\it complete }
 answer of the question about the {\it exact quantization } of the electrically
 charged topological solitons of the dyonic integrable models 
(\ref{1.1})  and  (\ref{1.3}), as well as whether they can be described as
representations of the {\it affine quantum group } $U_{q}(A_n^{(1)})$ requires
further investigations.

An important feature  of all $A_n^{(1)}$ abelian affine Toda models ($n
>1$) is known to be  that their action (as well as the Hamiltonian ) becomes complex 
for imaginary coupling
constants.  The corresponding soliton solutions are also complex functions but their
 energy and
momentum are real \cite{holl}.  The only exception is the Sine-Gordon model (i.e. $A_1^{(1)}$) 
whose action,
Hamiltonian, solitons and breathers are all real.  A similar phenomena takes place for the 
 $A_n^{(1)}$ dyonic IM ($n>1$).  There exist only two exceptional cases with real action,
Hamiltonian, etc. The first is the $A_1^{(1)}$ vector (or axial ) model (i.e. all $\varphi_i =0$)
which is known as the Complex Sine-Gordon (or Lund-Regge) model \cite{lund1},\cite{devega}.
  However it is not of dyonic
type since its solitons are either charged but nontopological or neutral topological in the T-dual
picture.  The $A_2^{(1)}$-vector model represents the unique member of the $A_n^{(1)}$ 
family
 of models (\ref{1.1}) and (\ref{1.3}) that has real actions, soliton solutions, etc.  After an
appropriate change of variables, $ A = e^{i\b_0 \theta  } \sinh (\b_0 r), \;\;  B = 
-e^{-i\b_0\theta  } \sinh (\b_0 r), \;\; c= e^{i\b_0 \varphi }$ 
we end up with the following real
${\cal L}_v$
\br
{\cal L}_v = \pa \varphi \bar \pa \varphi + \pa r \bar \pa r + \tanh ^2 (\b_0 r) \pa \theta \bar
\pa \theta + {{2m^2}\o {\b_0^2}}\sinh (\b_0 r) \cos \b_0 (\theta + 2 \varphi )
\nonu
\er
The properties of its soliton solutions can be extracted from eqns. (\ref{5.14}).

It is important to note that the $U(1)$ charged topological solitons {\it do not
exhaust} all the {\it particle-like finite energy } solutions of the IM's 
(\ref{1.1})  and  (\ref{1.3}).  The dyonic IM's in consideration admit also $d$ ``
species'' ($d=1, 2, \cdots n-1$) of neutral 1-solitons that turns out to coincide
with the $A_{n-1}^{(1)}$-abelian affine Toda solitons \cite{olive1},\cite{liao}.  The {\it complete
list } of solutions also includes {\it three} type of breathers representing all the
possible {\it bound states } of the charged and neutral 1-soliton and
anti-solitons (i.e. ``dyon-dyon'', ``dyon -monopole''  condensates, etc.).  The 
explicit construction of all these solutions  and the discussion of their
 spectrum are presented in our work \cite{ggsz2}.

Sect. 8 contains preliminary discussion of the S- and T- duality 
properties of the considered
integrable models.


\sect{$A_n^{(1)}$-Dyonic Model}
{\bf 2.1 Zero curvature and graded structures.} The {\it integrability} of
 the axial $A_n^{(1)}$-model is a consequence of the
fact that its equations of motion 
\br
&&\partial \left( \frac{\overline{\partial }{\psi }
e^{-\b \varphi _{1}}
}{\Delta }\right) +\b^2\left( \frac{n+1}{2n}\right) \frac{{\psi }
e^{-2\b \varphi _{1}}\partial {\chi }\overline{\partial }{
\psi }}{\Delta ^{2}}+ m^{2}e^{\b \varphi _{n-1} }{\psi }=0,
\nonu \\
&&\overline{\partial }\left( \frac{\partial {\chi }e^{-\b \varphi _{1}}
}{\Delta }\right) +\b^2\left( \frac{n+1}{2n}\right) \frac{{\chi }
e^{-2\b \varphi _{1}}\partial {\chi }\overline{\partial }{
\psi }}{\Delta ^{2}}+m ^{2}e^{\b \varphi _{n-1} }{\chi }=0 
\nonu \\ 
&&\pa \bar \pa \varphi_l + \b ({{n-l}\o {n}}){{\pa \chi \bar \pa \psi
e^{-\b \varphi_1}}\o {\Delta^2}} 
 - {{m^2}\o \b} e^{\b (\varphi_{l+1} +\varphi_{l-1} -2 \varphi_{l}) }+
\nonu\\
&&+ {{m^2}\o \b} \(e^{\b (\varphi_1 +\varphi_{n-1})} + \b^2 {{l\o n}}\psi \chi
 e^{\b \varphi_{n-1}} \) =0,  
\label{2.1}
\er
$l=1, \cdots , n-1$, can be written as zero curvature condition
\be
[\pa + {\cal A}, \bar \pa + \bar {\cal A}] =0. \label{2.2}
\ee
The Lax connections ${\cal A}$ and $\bar {\cal A}$ (i.e. ${\cal A_{\mu}}$)
lies in the   $A_n^{(1)}$ centerless affine Kac-Moody algebra (loop algebra) \footnote{ all the algebraic definitions
concerning $A_n^{(1)}$ are as in refs.\cite{olive1},\cite {ggsz2}} and are given by 
\br
{\cal A}&=&\b^2 \frac{\psi \partial \chi}{2\Delta}e^{-\b \varphi_1} 
{{\lambda_{1}\cdot H^{(0)}}\o {\l_1^2}}+{\b \o 2}\sum_{j=1}^{n-1}\partial
\varphi_{j}h_{j+1}^{(0)}
+\b \frac{\partial
\chi}{\Delta}e^{-{1\o 2}\b \varphi_1}E_{-\alpha_{1}}^{(0)} \nonu \\
&+&
{m }\( \sum_{j=1}^{n-1}
e^{-{1\o 2}\b  (2\varphi_{j}-\varphi_{j-1}-\varphi_{j+1})} E_{-\alpha_{j+1}}^{(0)}
+ \b \psi e^{{1\o 2}\b \varphi_{n-1}} E_{\a_1+\a_2 +\cdots \a_n}^{(-1)}
+   e^{{1\o 2}\b (\varphi_1 + \varphi_{n-1})}E_{\a_2 +\cdots \a_n}^{(-1)}\)
\nonumber
\\
\bar{{\cal A}}&=&-\b^2 \frac{\chi \bar{\partial}\psi}{2\Delta}
e^{-\b \varphi_1 } 
{{\lambda_{1}\cdot H^{(0)}}\o {\l_1^2}}-{\b \o 2}\sum_{j=1}^{n-1}\bar{\partial}
\varphi_{j}h_{j+1}^{(0)}
-\b \frac{\bar{\partial}
\psi}{\Delta}e^{-{1\o 2}\b \varphi_1}E_{\alpha_{1}}^{(0)}\nonu \\
&-& {m}\(
\sum_{j=1}^{n-1}
e^{-{1\o 2}\b (2\varphi_{j} -\varphi_{j-1}-\varphi_{j+1}  )} E_{\alpha_{j+1}}^{(0)}
+ \b \chi e^{{1\o 2} \b \varphi_{n-1}} E_{-(\a_1+\a_2 +\cdots \a_n)}^{(1)}
+   e^{{1\o 2}\b (\varphi_1 + \varphi_{n-1})}E_{-(\a_2 +\cdots \a_n)}^{(1)} \)
\nonu \\
\label{2.3}
\er
In the above particular choice  for the connection ${\cal A}_{\mu }$ is
hidden an interesting  and rich {\it algebraic structure} known as the
Leznov-Saveliev method \cite{leznov} for constructing 2-D integrable models.  In
order to give an idea of how it works in our case of the singular non-abelian
affine Toda (dyonic) models, we transform ${\cal A}$ and $\bar {\cal A}$ by an
appropriate gauge transformation (dressing)
\br
  {\cal A}_{\mu}^W = W {\cal A}_{\mu}W^{-1} + W\pa _{\mu}W^{-1}, \quad W^{-1} = 
  e^{{1\o 2}\sum_{i=1}^{n-1}\b \varphi_i h_{i+1}^{(0)}} e^{\b \psi
  E_{\a_1}^{(0)}} e^{{1\o 2}\b R {{\l_1 \cdot H}\o {\l_1^2}}}
 \nonu 
 \er
 that leaves eqs. (\ref{2.1}) unchanged and transforms the Lax connection into
 the following suggestive form
 \be
  {\cal A}^W = D^{-1} \pa D + \eps_-, \quad   {\bar {\cal A}}^W=
 -D^{-1} \eps_+D
\label{2.4}
\ee
where 
\be
\eps_{\pm} = {{m}}\(\sum_{l=2}^{n} E_{\pm \a_l}^{(0)} +  
 E_{\mp (\a_2 + \cdots +\a_n)}^{(\pm 1)}\)
\label{2.5}
\ee
The group element $D \in SL(2) \otimes U(1)^{(n-1)}$   is parametrized as follows
\br
D_a &=& e^{{\b R\o {2\l_1^2}}\l_1 \cdot H^{(0)}}e^{\b \chi E_{-\a_1}^{(0)}}
 e^{\sum_{l=1}^{n-1}\b \varphi_{l}h_{l+1}^{(0)}}
 e^{\b \psi E_{\a_1}^{(0)}}e^{{\b R\o {2\l_1^2}}\l_1 \cdot H^{(0)}}\nonu \\
 &=& 
 e^{\b \tilde \chi E_{-\a_1}^{(0)}}
 e^{{\b R\o {\l_1^2}}\l_1 \cdot H^{(0)} + \sum_{l=1}^{n-1}\b \varphi_{l}h_{l+1}^{(0)}}
 e^{\b \tilde \psi E_{\a_1}^{(0)}}
 \label{2.6}
 \er
 As explained in refs. \cite{annals},\cite{ggsz2} the $\eps_{\pm}$-invariant
 subgroup $g_0^0 \in U(1)$, $[\eps_{\pm},g_0^0] = 0$, is spanned by $\lambda_1
 \cdot H^{(0)}$, i.e.
 $ g_0^0 = e^{{R \o {\l_1^2}}\lambda_1 \cdot H^{(0)}} $, 
  where $\lambda_l, l= 1, \cdots n$ are the  fundamental  weights of $A_n$,
   $\l_1^2 = {{n}\o {n+1}}$. 
   The appearence of the
  nonlocal field $R$ defined by 
  \be
  \pa R = \b {{\psi \pa \chi }\o \Delta }e^{-\b \varphi_1},
 \;\; \;\;\; 
\bar \pa R =  \b {{\chi \bar \pa \psi }\o \Delta }e^{-\b \varphi_1}
\label{2.7}
\ee
is a consequence of the subsidiary constraint 
\br
J = Tr (D^{-1}\pa D \lambda_1 \cdot H^{(0)}) =0, \quad  \quad
\bar J =Tr (\bar \pa D D^{-1}\lambda_1 \cdot H^{(0)}) = 0
\nonu 
\er
imposed on the zero grade  conserved currents ($\pa \bar J = \bar \pa J =
0$) of the original {\it two loop } gauged 
$H_{-}\backslash A_n^{(1)}\otimes A_n^{(1)}/ H_{+}$ WZW model\cite{ggsz2},\cite{aratyn}. 
 We remind the algebraic recipe 
for constructing generic integrable models(see for example 
\cite{olive1},\cite{babelon},\cite{luiz},\cite{ggsz2}):  Given an (finite or infinite
dimensional ) algebra $\lie_n^{(a)}$.  
Introduce a graded structure $(\lie_n^{(a)}, Q)$ by means of the grading
operator $Q$, 
\br
[Q, \lie_l] =l\lie_l, \quad \lie_n^{(a)} =  \bigoplus \lie_l, \quad 
[\lie_l, \lie_k] \in \lie_{l+k}\quad
l,k=0, \pm 1, \cdots
\nonu
\er
A family of grade one integrable models $\{ \lie_n^{(a)}, Q, \eps_{\pm},
g_0^f \subset \lie_0 /\lie_0^0 \}$ is defined by:
\begin{itemize}
\item an appropriate choice of grade one constant elements $\eps_{\pm} \in
\lie_{\pm 1}$.
\item The $0$-grade group element $D = exp {(\lie_0 )}$ contains all
fields $R(z, \bar z), \varphi_l (z, \bar z), \psi(z, \bar z), 
\chi (z, \bar z),$ etc.  appearing in (\ref{1.1}), (\ref{2.1}), (\ref{2.3}).
\item When $\lie_0$ has an invariant  subspace $\lie_0^0 \subset \lie_0$, such that
$[\eps_{\pm}, \lie_0^0 ] =0$, one can consider the subfamily of {\it ``singular''
integrable models } by imposing the subsidiary constraints
\be 
Tr(D^{-1}\pa D \lie _0^0)= Tr(\bar \pa D D^{-1} \lie _0^0) =0
\label{bh} 
\ee
allowing to  eliminate the degree of freedom associated to $\lie _0^0$.
\item Finaly, with $\eps_{\pm}$, $D = exp {(\lie_0 )}$ and the subsidiary
condition (\ref{bh}) we can construct the desired Lax connections ${\cal A}, 
\bar {\cal A}$ according to eqs. (\ref{2.4}).
\end{itemize}

The answer to the questions concerning the derivation of this recipe, its
equivalence to the Hamiltonian reduction (Drinfeld-Sokolov) procedure and the
two loop gauged $H_- \backslash G_n^{(a)} \otimes G_n^{(a)}/H_+$ WZW models as
well as the classification of the {\it grade one } integrable models can be
found in refs. \cite{ggsz2}.  The $A_n^{(1)}$-{\it dyonic model}
(\ref{1.1}) in consideration corresponds to the following specific choice of
the grading operator 
\br
Q = n\hat d + \sum_{l=2}^{n} \lambda_l \cdot H^{(0)}, \quad [\hat d,
E_\a^{(p)}] = p E_\a^{(p)}, \quad [\hat d,
h_i^{(p)}] = p h_i^{(p)}, \;\; \;\;\; p= 0,  \pm 1, \cdots
\nonu
\er
The grade one constant generators $\eps_{\pm} $ are those  given in eq.
(\ref{2.5}) and $ \lie_0^0 = \lambda_1 \cdot H^{(0)}$. 

{\bf 2.2 $A_n^{(1)}$- Vector Model.}  As we have mentioned, there exist two inequivalent
 ways (axial and vector)  of gauge fixing the local $U(1)$ symmetry
generated  by the currents $(J_{\l_1 \cdot H}, \bar J_{\l_1 \cdot H}) \in \lie_0^0$. 
 The details concerning the derivation of the axial and
  vector Lagrangeans, (\ref{1.1}) and (\ref{1.3})
   from the ungauged one (\ref{1.4})  are
 presented in our recent paper \cite{ggsz1}.  The problem we  address here is about 
 the nonlocal change of the axial variables $\psi, \chi,
 \varphi_i $ into the vector ones $A, B, c_i$. Observe that  both, the axial model eqns.  (\ref{2.1}) 
 as well as the vector model
 equations of motion 
 \br
 & \pa \bar \pa   c_1 & = m^2 \({{c_2}\o {c_1}} - A c_1^2 c_2 \cdots c_{n-1}\), 
 \quad \quad 
  \pa \bar \pa c_{n-1} =  m^2 \({{B}\o {c_1 \cdots c_{n-1}^2}} - {{c_{n-1}}\o
{c_{n-2}}}\),\nonu \\
& \pa \bar \pa   c_k  & = m^2 \( {{c_{k+1}}\o {c_{k}}}- {{c_{k}}\o {c_{k-1}}}\) , 
\quad \quad k=2,3, \cdots n-2 \nonu \\
& \pa \( {{\bar  \pa A}\o {1- AB}}\) &= {{A\bar \pa A \pa B}\o {(1- AB)^2}} +
{{m^2}\o {c_1 \cdots c_{n-1}^2}} \nonu \\
& \bar \pa \( {{  \pa B}\o {1- AB}}\) &= {{B\bar \pa A \pa B}\o {(1- AB)^2}} +
 m^2{c_1^2 c_2 \cdots c_{n-1}}
 \label{5.1}
 \er
 can be written in a compact form    
\be 
\bar \pa (D^{-1} \pa D) + [ {\eps_-}, D^{-1}  {\eps_+} D] =0, 
\quad \pa (\bar \pa D D^{-1} ) - [ {\eps_+}, D {\eps_-} D^{-1}] =0
\label{2.9}
\ee
For the axial case we take $D = D_a\in g_0$ in the form (\ref{2.6}),
 or equivalently in the following matrix representation 
 \br
 D_a = \twomat{\tilde d_2}{0}{0}{\tilde d_{n-1}},\quad \tilde d_{n-1} = 
 diag (e^{\b(\varphi_2 - \varphi_1 - {{R\o n})}}, \cdots ,
  e^{-\b(  \varphi_{n-1} + {{R\o n}})} ) \nonu
 \er
 \br
  \tilde d_2 = 
  \twomat{e^{\b R}}{\b \psi e^{\b {{n-1}\o {2n}}R}}
  {\b \chi e^{\b {{n-1}\o {2n}}R}}{
  e^{\b (\varphi_1 - {R\o n})}(1+ \b^2 \psi \chi e^{-\b \varphi_1})}
 \label{5.4}
 \er
Eliminating further the field $R$ according to eqns. (\ref{2.7}) we derive 
eqns. (\ref{2.1}) from (\ref{2.9}) and (\ref{2.7}).  The
parametrization  of $D \in SU(2)\otimes U(1)^{n-1}$ appropriate for the vector case is 
\br
 D_v = \twomat{d_2}{0}{0}{d_{n-1}}, \quad
  \quad d_{n-1} = diag (c_1, \cdots c_{n-1}) \nonu
 \er
 \br
  d_2 = \twomat{A}{u}{{{AB-1}\o{uc_1c_2 \cdots c_{n-1}}}}{{{B}\o {c_1c_2 \cdots
  c_{n-1}}}}
 \label{5.2}
 \er
 where the nonlocal field $u$ ( the vector model analog of $R$) is defined by the 
 following first order equations 
 \br
 \pa ln \(u c_1 \cdots c_{n-1} \) = - {{A \pa B}\o {1-AB}}, \quad \bar \pa ln u = - {{B \bar \pa A}\o {1-AB}}
\label{vecu}
\er
Starting from eqns. (\ref{2.9}) with $D= D_v$ and taking into account (\ref{vecu}) 
the result now is the vector model equations 
(\ref{5.1}).  It is then clear that comparing axial and vector parametrizations of 
$D$, i.e. $D_a = D_v$ we arrive at the desired axial-vector
change of variables 
\br
 e^{\b R} &=& A, \quad e^{-\b \varphi_k} = c_k c_{k+1} \cdots c_{n-1}A^{{{n-k}\o
 {n}}}, \quad k=1, \cdots , n-1 \nonu \\
 \b \psi &=& uA^{- {{n-1} \o {2n}}}, \quad \b \chi = {{AB-1}\o {uc_1 c_2 \cdots
 c_{n-1}}}A^{- {{n-1} \o {2n}}}, \quad 
 \label{2.23}
 \er
  Equivalently the reverse vector-axial transformations are given by
\br
c_k = e^{\b (\varphi_{k+1} - \varphi_{k} - {R\o n})}, \quad A&=&e^{\b R},
\quad B=e^{-\b R}( 1+ \b^2 \psi \chi e^{-\b \varphi_1}) \nonu \\
\tilde u ^2 = {{\psi }\o {\chi }}, && \tilde u = u \({{c_1 c_2 \cdots c_{n-1}}\o {AB-1}}\)^{1\o 2} 
\label{3.10}
\er
One can easely check that inserting (\ref{2.23}) in eqns. (\ref{2.1}) gives the 
vector model equations (\ref{5.1}).  It is worthwhile to
mention that axial gauge fixing corresponds to the elimination of the field $R$ from $\lie_0^0$ constraints, 
$J_{\l_1 \cdot H} = \bar J_{\l_1 \cdot H} =0$,
i.e. eqns. (\ref{2.7}) in this case.  In this language, the vector gauge fixing  is equivalent
 to the elimination of another field ${{\psi}\o
{\chi }} = \tilde u^2$ from the same constraint equations.

Performing the above change of variables in ${\cal L}_v$ we find the following relation 
including the so called generating function
\cite{ggsz1} ${\cal F}$:
\be
{\cal L}_v = {\cal L}_a + {{d {\cal F}}\o {dt}}, \quad \quad \quad 
\label{func}{{d {\cal F}}\o {dt}} = {{1\o {2\b}}}\( \pa R \bar \pa ln ({{\chi}\o {\psi }}) 
- \bar \pa R  \pa ln ({{\chi}\o {\psi }}) \)
\ee
It is not surprising
that the same relation (\ref{func}) appears as a result of abelian 
T-duality transformation between the axial and vector IM's in
consideration (see Sect 3 of ref. \cite{ggsz1}).  Denote by 
$\theta = -{i\o {2 \b_0}}ln \( {{\chi }\o {\psi}}\) $ (for the axial model
(\ref{1.1})) and $\tilde \theta = {{2}\o {i \b_0}}ln A = 2 R$ 
(for the vector model ) the corresponding isometric coordinates
\footnote {$\theta \rightarrow \theta + a$ and 
$\tilde \theta \rightarrow \tilde \theta + \tilde a$ are symmetries of ${\cal L}_a$ and 
${\cal L}_v$, that coincides with their global $U(1)$ symmetry} 
 and by $\Pi_{\theta}$ and $\Pi_{\tilde \theta}$ their conjugate momenta. 
As it is well known \cite{ggsz1} , \cite{giveon} the following 
canonical transformation
\br
\Pi_{\tilde \theta} = - \pa _{x} \theta, \quad \Pi_{ \theta} = - \pa _{x} \tilde \theta
\label{cann}
\er
(and all remaining $\Pi_{\varphi_i}, \varphi_i$ unchanged) acts as 
T-duality transformation   with generating
function ${\cal F}$\cite{giveon}.  Then, the relation  (\ref{func}) between vector
 and axial Lagrangeans is a simple consequence of the fact that both
Hamiltonians are equal, i.e. $H_a = H_v$.  Therefore the nonlocal 
change of variables (\ref{2.23}) represent an integrated form of the
T-duality transformations (\ref{cann}) accompanied by certain point 
transformations of the rest of the (nonisometric ) variables
$\varphi_k \rightarrow c_k = f_k(\varphi_i, R)$.

{\bf 2.3 Symmetries and Vacua.}  It is important to note
that both, the Noether symmetries of eqs. (\ref{2.1}) ( and ${\cal L}_{a}$ in
(\ref{1.1})), as well as the multiple zeros of the potential (\ref{1.1}) 
\br 
V_a &=& {1\o {\b^2}}\( Tr(\eps_+ D \eps_- D^{-1}) - {m^2}n  \)
\label{2.8}
\er
 are encoded in the {\it
algebraic data}: ($Q, \eps_{\pm}, \lie_0^0$).  Writing the field equations
(\ref{2.1}) in a compact form  (\ref{2.9}) and taking traces with $h_0^i$, 
we verify that all the elements $h_0^{i}$ of $\lie_0^0 \subset \lie_0$ ($[h_0^{i},
\eps_{\pm}] =0, \; \; i=1, \cdots , K_0=dim \;\; \lie_0^0$ ) generate {\it continuous
symmetries} of
 eqn. (\ref{2.9})
 \be
 D^{\pr} = e^{-i \b_0^2 \sum_{l=1}^{K_0} a^{l} h_0^l} D
 \label{2.10}
 \ee
 In our case $K_0=1$, $h_0^{(1)} = \lambda_1 \cdot H^{(0)}$ and the above
 transformation is a global $U(1)$ (electric) symmetry of (\ref{1.1}) mentioned
 in Sect. 1.  For {\it imaginary} coupling $\b = i\b_0 $ ,
  the $A_n^{(1)}$-{\it dyonic}
 Lagrangean (\ref{1.1}) ( as well as the potential (\ref{2.8})) are invariant
  under the following {\it discrete group} transformations 
 \be
 \varphi^{\pr}_l = \varphi_l + {{2 \pi}\o \b_0}{l\o n} N, \;\;\;\; l= 1, \cdots
 ,n-1,
 \quad \psi^{\pr} = e^{i\pi ({N\o n} + s_1)}\psi, \quad 
 \chi^{\pr} = e^{i\pi ({N\o n} + s_2)}\chi
 \label{2.11}
 \ee
 for $N$ an arbitrary integer and $s_a, a=1,2$ are both even ( odd) integers (
  i.e., $s_1 + s_2 = 2S^{\pr}, s_1 - s_2 =2L^{\pr}$). It is  also invariant under CP transformations ($P: x \rightarrow -x$)
  \br
  \varphi_l^{\pr \pr } = \varphi_l, \quad \psi^{\pr \pr } = \chi, \quad \chi^{\pr \pr } = \psi
  \label{cptrans}
  \er
  It is convenient to 
   parametrize $\psi$ and $\chi$ as
  \br
  \psi = {1\o {\b_0}}e^{i\b_0 ({1\o 2}\varphi_1  -\theta ) }\sqrt {{{2n}\o {n+1}}} \sinh (\b_0 r),
  \quad 
  \chi ={1\o {\b_0}} e^{i\b_0 ({1\o 2}\varphi_1 +\theta ) }\sqrt {{{2n}\o {n+1}}} \sinh (\b_0 r)
  \nonu
  \er
  in terms of one noncompact  ($r$) and two compact ($\varphi_1, \theta$)
  fields following the tradition of the 3-D {\it black string } constructions
  \footnote{our $n=2$ ``free'' ${\cal L}$ (i.e. $V=0$) after an appropriate
  field redefinitions (reflecting another $U(1)$ gauge fixing ) and with the counterterms
  (\ref{6.13}) and (\ref{6.13a}) included coincides with the euclidean 3-D black string of ref. 
    \cite{horne}.}. 
    The $\psi, \chi$ discrete transformations (\ref{2.11}) and (\ref{cptrans}) then take the form
    \br
    r^{\pr} &=& r, \quad \theta^{\pr} = \theta + {{\pi }\o {\b_0}}L^{\pr}, \nonu \\
     r^{\pr \pr } &=& r, \quad \theta^{\pr \pr } = -\theta + {{2 \pi }\o {\b_0}}L^{\pr \pr }
     \label{rtheta}
     \er
     with $L^{\pr}$ and $L^{\pr \pr}$ arbitrary integers.
    As a consequence of (\ref{2.11}) and (\ref{cptrans}) 
    our
  dyonic model (\ref{1.1}) possesses together with the  trivial classical
  vacuum solution $D=1, (N=0, \varphi_l= 0, \psi \chi =0,  \ln {{\psi }\o {\chi }} = const.,  
  V(D=1)=0)$  an infinite set
  of {\it new  distinct vacua}
  \be
  \varphi_l^{(N)} = {{2\pi }\o \b_0}{l \o n}N, \quad r^{(0)} =0, \quad \theta
  ^{(L)} = -{{i}\o {2\b_0}}ln ({{\chi^{(L)}} \o {\psi^{(L)}}}) = 
  {{\pi L}\o {2\b_0}}, \quad L= 2 L^{\pr \pr }-L^{\pr}
  \label{2.12}
  \ee
  This provides a new set of allowed boundary conditions for the fields at $x \rightarrow \pm \infty$:
  \br
  \varphi_l^{(N)} (\pm \infty ) = {{2\pi}\o {\b_0}}{{lN_{\pm}}\o {n}}, \quad r(\pm \infty ) = 0, 
  \quad \theta^{(L)}(\pm \infty )= {{\pi
  L_{\pm}}\o {2\b_0}}\nonu 
  \er
  (i.e. $\psi (\pm \infty ) =\chi (\pm \infty )=0$, but ${{\psi}\o {\chi }}(\pm \infty ) \neq 0$).
The multiply degenerate vacuum  of our
  axial integrable model (\ref{1.1})  is an indication   that it admits {\it
  finite energy topological solitons} solutions  interpolating between two
  different vacua  ($N_{\pm}, L_{\pm}$): $ j_{\varphi} = N_+ -  N_- (mod \; n)$, $
  j_{\theta} = L_+ - L_-$.   Due to the $U(1)$ symmetry (\ref{2.10}) ($K_0=1$)
  such solutions might also carry nontrivial {\it electric charge}:
  \be
  Q_{el}^{ax} = \int_{-\infty}^{\infty} dx J_{el}^{0, ax} = 
  2\b_0 \int_{-\infty}^{\infty} dx  \pa_x R, \quad J_{el}^{\mu , ax} =
  2\b_0 \eps^{\mu \nu } \pa _{\nu} R
  \label{2.14}
  \ee
  where $R$ is defined in (\ref{2.7}).  The topological current
  \be
  J^{\mu , ax}_{\theta} = { {\b_0}}\eps ^{\mu \nu } \pa _{\nu} \theta ( = {1\o
  {2i}}\eps^{\mu \nu}\pa_{\nu} ln ({{\chi} \o {\psi}})), \quad 
   Q_{\theta}^{ax} = {{\pi j_{\theta}}\o {2}}, \quad j_{\theta}= 0, \pm 1, \cdots 
  \label{jota}
  \ee
   appears to be T-dual to the electric current $J^{\mu ,vec}_{el} $ (see
  Sect. 2.5 and ref. \cite{ggsz1}).  It plays the role of electric current in the
  T-dual model (\ref{1.3}) that generates the  global $U(1)$ transformation: 
  $A^{\pr} = e^{ia\b_0^2}A, B^{\pr} = e^{-ia\b_0^2}B, c_k^{\pr}
   =
  e^{-i{a \o {n}}\b_0^2}c_k$.  Similarly  to the free case, ($V=0$), \cite{giveon} the
  currents  $J^{\mu ,ax}_{el},  J^{\mu ,ax}_{\theta}$ ( and their conjugate
  ``coordinates'' $\theta$ and $2 R\equiv \tilde \theta $) form a canonical pairs
  \br
  \{{1\o {\b_0}}J_{el}^{0,ax}(x, t_0), \theta (y, t_0)\} = -
  \{\tilde \theta (x, t_0), {{1\o {\b_0 }}}J_{\theta }^{0,ax}(y, t_0)\} = \d(x-y)  
 \nonu
 \er
 but $\{Q_{el}^{ax}, Q_{\theta}^{ax}\} =0$.   Note that  the other topological currents
 \br
 J^{\mu}_k = {{2n }\o {\b_0}} \eps^{\mu \nu} \pa _{\nu} \varphi_k,
  \quad  Q_k^{top} = \int _{-\infty}^{\infty}J^0_{k} dx = k Q_{mag}, \quad Q_{mag} = {{4\pi} \o
  {\b_0^2}}j_{\varphi}
  \label{carga}
  \er
  ( $Q_{mag}$ stands for the magnetic charge) 
   {\it are not } T-duals to the corresponding currents $\tilde
 J^{\mu}_{k} $ (\ref{5.21}) of the vector model (\ref{1.3}).

   The origin of the {\it discrete } symmetry (\ref{2.11}) is
  in  the following {\it continuous } symmetry of the $V=0$ free   model (\ref{1.1})
 \br
 J^{\mu}_{\varphi_1}&:& 
 \varphi_1^{\pr} = \varphi_1 + a_1, \quad \psi^{\pr} = e^{{i\b_0 {a_1 }\o 2}}\psi
 , \quad \chi^{\pr} = e^{{i\b_0 {a_1 }\o 2}}\chi
 \nonu\\
 J^{\mu}_{\varphi_k}&:& \varphi_k^{\pr} = \varphi_k + a_k, \quad \psi^{\pr} =
 \psi, \quad \chi^{\pr} = \chi,\;\;\;\; k\neq 1 
 \nonu \\
 J^{\mu}_{el} &:& \varphi_l^{\pr} =\varphi_l, \quad  \psi^{\pr} = e^{ia \b_0^2}\psi, \quad  \chi^{\pr} =
  e^{-ia \b_0^2}\chi, \;\;\;\;\; l=1,2,\cdot,n-1
\nonu
\er
 broken to (\ref{2.12}) when $V \neq 0$ is added.  It is  important to mention
 an interesting relation between the {\it two} $U(1)$ currents
 $J^{\mu}_{\varphi_1}, J^{\mu}_{el}$ and the topological current
  $I^{\mu}_{\varphi_1} = \eps ^{\mu \nu} \pa_{\nu}\varphi_1 $ which, say for
  $n=2$ ,reads 
  \be
  J^{\mu}_{el} = 2 \eps^{\mu \nu} J_{\nu}^{\varphi_1} - 2 I^{\mu }_{\varphi_1}
  \label{2.15}
  \ee
  {\bf 2.4 Topological $\theta$-term.}
  Although eqn. (\ref{2.15}) takes place in the $V=0$ model only,  the fact
  that {\it the electric charge gets contributions from the topological charge }
  $Q_1^{top}$ persists in the general case where $V\neq 0$ (\ref{1.1}) as we
  shall show on the example of the 1-soliton solutions in Sect. 3.  As one
  can expect  the {\it dyonic properties} of our model 
  (\ref{1.1}) are consequences of  the CPT breaking term $ {{\eps ^{\mu \nu } \pa_{\mu}\psi 
   \pa_{\nu}\chi }\o {\Delta}}e^{-i\b_0 \varphi_1} = \d {\cal L}_{CPT}$
   which, by
   the following change of variables, $\psi_0 = e^{-i{\b_0 \o 2}\varphi_1} \psi,
   \; \chi_0 = e^{-i{\b_0 \o 2}\varphi_1} \chi$ reads, 
   \br
   \d {\cal L}_{CPT} = {{\eps^{\mu \nu}}\o {\Delta}} \{ \pa _{\mu} \psi_0
   \pa_{\nu} \chi_0 - {{in}\o {\b_0{(n+1)}}} \pa _{\mu} \varphi_1 \pa _{\nu} ln
   ({{\chi_0}\o {\psi_0}})\} + {{i n }\o {\b_0 {(n+1)}}}
   \eps^{\mu \nu}\pa _{\mu} \varphi_1\pa _{\nu} ln
   ({{\chi_0}\o {\psi_0}})
   \nonu
   \er
   The {\it pure topological} term 
   \br
   {{in} \o {\b_0 {(n+1)}}}  \pa ^{\nu} \( \eps_{\mu \nu } 
   ln ({{\chi_0}\o {\psi_0}})
   \pa ^{\mu} \varphi_1 \) \equiv {{2n}\o {n+1}}  \eps^{\mu \nu } \pa
   _{\mu} \varphi_1 \pa _{\nu} \theta 
   \nonu 
   \er
(being a total derivative) 
  does not contribute to the equations of motion (\ref{2.1}),
but makes evident the top-charge contribution to the electric current
 (\ref{2.14}),
$ \d J^{\mu}_{el} =   {{2n}\o {n+1}} \b_0 \eps ^{\mu \nu} \pa_{\nu}
\varphi_1 $.  We further observe that the specific value of
 the coefficient
multiplying the topological current is irrelevant since one
 can make it arbitrary by
adding  to the original Lagrangean (\ref{1.1}) certain ``$\theta$-type ''
 topological
terms  \footnote{We  restrict ourselves to consider $\d {\cal L}_{top}^{a}$ in
 a particular form (\ref{2.16}), although one can include more
$\theta$-terms, say, $\nu_{ij} \pa_{\mu} \varphi_i \pa_{\nu} \varphi_j \eps^{\mu \nu} + 
\nu_{i0}\eps^{\mu \nu} \pa_{\mu} \varphi_i \pa_{\nu} ln \Delta + \cdots$.  
The only contribution to $Q_{el}$ comes from (\ref{2.16}).}
\be
\delta {\cal L}_{top}^{a} = i {{\b_0}\o {8\pi^2}} \sum_{k=1}^{n-1} 
\nu_k \eps ^{\mu \nu } \pa_{\mu} \varphi_k
\pa_{\nu} ln ( {{\chi }\o {\psi}})
\label{2.16}
\ee
where $\nu_k$ are arbitrary real  constants.  As a result, the improved
 electric
current ( calculated from ${\cal L}_{impr} = {\cal L} + \delta {\cal
L}_{top}$)
has the form
\be
J^{\mu , ax}_{el, impr} =  2\b_0 \eps ^{\mu \nu } \pa_{\nu} R - {{\b_0^3}\o
{4\pi^2}}\eps ^{\mu \nu }
 \sum_{k=1}^{n-1} \nu_k \pa_{\nu} \varphi_k
\label{jelimpr}
\ee
Therefore we find  that $Q_{el}^{ax}$ is shifted by $\nu j_{\varphi}$:
\be
\int_{-\infty}^{\infty} J^{0, ax}_{el, impr} dx = Q_{el}^{ax} - {{\nu \b_0^2}\o
{2\pi}}j_{\varphi}, \quad \nu=  {1\o n}\sum_{k=1}^{n-1} k \nu_k, \quad
j_{\varphi} = 0, \pm 1, \cdots , \pm (n-1)
\label{cargael}
\ee
Similar phenomenon takes place in the open string with one compactified
 dimension (say $X_{25} \equiv \theta = {1\o {2 \b}} ln {{\psi}\o
{\chi}}$) and boundary (Chan-Paton) term included ( see Sect 8.6 of ref. \cite{polch}).  The string 
momenta $P_{25}$ gets contribution from the boundary (Wilson line)
term $\Delta S_{string}= i \oint A_{25}(X_i) dX_{25}$.  Such similarity 
is not occasional, since in our case the electric current $J_{el}^{ax}$
coincides with the momenta $\Pi_{\theta}$ conjugate to the ``field 
 coordinate '' $\theta$, which for 1-soliton solutions (\ref{4.9}) turns
out to be periodic.  Then the $\theta$-term (\ref{2.16}) we have added 
to the original action (\ref{1.1}) can be rewritten as the Wilson
line of certain background gauge field $A_a(X_k) (a=1, 2, \cdots n+1$ 
and  $X_k = \{ \varphi_i, \theta , ln \Delta \}, k=1,2, \cdots n+1)$:
\br
\d S_{top}^{a} = \int d^2z \d {\cal L}_{top}^{a} = 
\({{\b}\o {4 \pi}}\)^2 \oint dt \nu_i \varphi_i \pa_t \theta = \oint d \theta
A_{\theta}(X_{k})
\label{topint}
\er
i.e. $A_{\theta} = \( {{\b}\o {4 \pi}}\)^2 \sum_{i=1}^{n-1} \nu_i \varphi_i $ and 
$ A_a =0$ for $a \neq \theta $.  Therefore our specific
$U(1)^{n-1}$ linear gauge potential $A_a(X)$ corresponds to constant
 `` electromagnetic field '' $F_{ab}= \pa_a A_b - \pa_b A_a$, i.e.
$F_{\theta i} =  \({{\b}\o {4 \pi}}\)^2 \nu_i$ and all the other components vanish.  

{\bf 2.5 Axial vs. Vector $\theta$-terms.}
The shift of the electric charge $Q_{el}^{ax} \rightarrow 
Q_{el}^{ax} - {{\nu \b_0^2}\o {2 \pi }}j_{\varphi}$ induced by 
the topological term
(\ref{2.16}) together with the fact that $Q_{\theta}^{ax}$ 
is equal to the vector model electric charge $ Q_{el}^{v}$ (i.e.
$Q_{el}^{ax}, Q_{\theta}^{ax} \rightarrow  Q_{\tilde  \theta}^{v}, Q_{el}^{v}$
by T-duality) addresses the question about the changes in $Q_{\theta}^{ax}$ 
caused by (\ref{2.16}).  Its answer requires the explicit form
of the vector model  $\theta$-term, that corresponds to the axial one (\ref{2.16})
 via T-duality transformation.  Applying the
axial-vector change of variables (\ref{2.23}) in eqn. (\ref{2.16}) results in a 
{\it nonlocal} $\theta$-term (including $\tilde u$) for the vector
model.  This is an indication that one should consider new canonical 
transformations (more general than (\ref{cann})) in order to  have {\it local}
$\theta$-terms for the vector model. Therefore we choose
\br
\tilde \theta_{impr} &=& \tilde \theta + \sum_{k=1}^{n-1} \a_k \varphi_k, \quad 
 \theta_{impr} =  \theta + \sum_{k=1}^{n-1} a_k \varphi_k +
 a_0ln\( 1 + \b^2 \psi \chi e^{-\b \varphi_{1}}\) \nonu \\
 \tilde \theta &=& 2 R, \quad \theta = {{1}\o {2 \b}}ln \( {{\chi }\o {\psi}}\)
 \label{tilde}
 \er
 as new isometric coordinates.  Next, we define the standard (T-duality) canonical 
 transformation (\ref{cann})  in terms of the new isometric coordinates $\tilde
 \theta_{impr}, \theta_{impr}$ and their conjugate momenta (together with the 
 point transformations in the first line of eqn. (\ref{3.10})). 
 In order to simplify the calculation we seek for the vector $\theta$-term in a
  form:
 \br
 \d {\cal L}_{top}^{v} = 
 \sum_{k=1}^{n-1} {{\g_k}\o {\b^2}}\eps^{\mu \nu } \pa_{\mu }
 ln \( c_k \cdots c_{n-1} \) \pa_{\nu} ln \({{A}\o
 {B}}\) 
 \label{varl}
 \er
 The consistency with the ``improved'' T-duality transformation  
 (with $\tilde \theta_{impr}, \theta_{impr}$) determines the unknown
 constants $\a_k, a_k, a_0$ and $\g_k$ as follows:
 \br
 \a_k &=& {{\nu_k \b^2}\o {4 \pi^2}}, \quad a_k = -\g_k, \quad  
 a_0 = {{1}\o {2n\b}}\sum_{k=1}^{n-1} \g_k (k-n) \nonu \\  \nu_k &=& 
 {{8\pi^2 n}\o {\b^2}} {{\g_k}\o {\sum_{k=1}^{n-1}\g_k (n-k)}}, \quad 
 {{\g_k}\o {\nu_k}} =  {{\g_l}\o {\nu_l}}= - {{a_0 \b^3}\o {4\pi^2}}, \;\; k,l= 1,2, \cdots , n-1
 \label{alpha}
 \er
As it is well known the $U(1)$- currents and the $\theta$ and 
$\tilde \theta$ topological currents of the  axial and vector models are related as follows:
\br
J_{el, impr}^{0, ax}  = \b_0 \pa_x \tilde \theta_{impr} = J_{\tilde \theta}^{0, vec}\quad 
J_{el, impr}^{0, vec}  =\b_0 \pa_x  \theta_{impr} = J_{ \theta}^{0, ax}
\label{jaxjvec}
\er
The corresponding charges calculated from the Lagrangeans with 
$\theta$-terms and by using the explicit form (\ref{tilde}) of $\tilde
\theta_{impr}$ ( and $\theta_{impr}$) indeed coincide,
\br
Q_{el, impr}^{ax} = Q_{\tilde \theta}^{vec} = Q_{el}^{ax} - 
{{\nu \b_0^2}\o {2 \pi}}j_{\varphi},  \quad
Q_{el, impr}^{vec} = Q_{ \theta}^{ax} = 
Q_{\theta} + 2 \pi \tilde \g j_{\varphi}, \label{qq}
\er
The new parameters $\tilde \g, \g_0, \nu_0$ and $\nu$ are defined by
\br
 \tilde \g = {1\o n} \sum_{k=1}^{n-1} k\g_k, \quad \g_0 = \sum_{k=1}^{n-1}\g_k, 
 \quad  \nu ={1\o n}\sum_{k=1}^{n-1}k\nu_k, \quad 
 \nu_0 = \sum_{k=1}^{n-1}\nu_k \nonu
\er
and satisfy the following relations 
\br
\nu_0 &=& \nu + {{8\pi^2}\o {\b^2}}, \quad \nu =  
{{8\pi^2}\o {\b^2}} {{\tilde \g }\o {\g_0 - \tilde \g }}, \quad \tilde \g = {{\g_0 \nu
}\o {\nu + {{8\pi^2}\o {\b^2}}}}
\label{qqqq}
\er
Eqns. (\ref{qqqq}) are direct consequence of the above definitions and eqns. (\ref{alpha}).  
Finally, the shifts in the charges due to the $\theta$-terms take the form:
\br
Q_{el, impr}^{ax} = Q_{el}^{ax} - {{\b^2 \nu }\o {2\pi }} j_{\varphi}, \quad 
Q_{\theta , impr}^{ax} = Q_{\theta} + {{2\pi \g_0 \nu}\o {\nu + {{8\pi^2}\o {\b^2}}}}, \quad 
Q_{\theta} = {\pi \o 2}j_{\theta}
\label{aaqq}
\er

{\bf 2.6 More discrete symmetries.}
 The ``translational type'' ($\varphi_l, \theta$)-symmetries (\ref{2.11}) that allows 
us to determine the {\it vacua lattice } ($\varphi_l^{(N)}, \theta^{(N)} $)
(\ref{2.12}) does not exhaust all the discrete symmetries of the $A_n^{(1)}$ dyonic 
Lagrangean (\ref{1.1}).   By analogy with the $S_n$ symmetries of the $A_n$-abelian
Toda models  one expects certain ($\varphi_l, \theta$)-``rotational''
symmetries to take place.
It is easy to verify that the vector  model  (\ref{1.3}) is invariant
under the following transformation 
\be
A^{\pr} = B, \quad  B^{\pr} = A, \quad c_k^{\pr} ={1\o {c_{n-k}}}, \;\; k=1, 2,
\cdots n-1
\label{2.18}
\ee
The derivation of their counterpart that leaves invariant the axial gauged
model (\ref{1.1}) is however far from obvious.  We start with the {\it
observation } that the eqs. (\ref{2.9}) as well as the potential (\ref{2.8})
remain invariant ($V(D^{\pr}) = V(D)$) under the  transformation $\Pi$:
$D^{\pr} = \Pi (D^{-1})$ with the following properties:
\be
\Pi (\eps_+) = \eps_-, \quad \Pi (E_{\a_1}) = E_{\Pi (\a_1)} = E_{\a_1}, \quad
\Pi^2 = 1
\label{2.19}
\ee
combined with the space reflection $P$: $P \pa = \bar  \pa$, $P^2 =1$.  The next
step is to explicitly construct $\Pi$ with the above properties in terms of the
elements $w_{k} = w_{\a_k}, (w_k \b =\b - (\a_k \cdot \b)\a_k )$ of the Weyl
group of $A_n^{(1)}$.  Define  a composite transformation ($ n \geq 3$),
\br
\Pi_0 ^{(n=2j)} &=& t^{j+2}_{j+1} t^{j+3}_{j}\cdots
t^{2j-1}_{4}t^{2j}_{3},\nonu \\
\Pi_0 ^{(n=2j-1)} &=& 
t^{j+1}_{j+1} t^{j+2}_{j}\cdots t^{2j-2}_{4}t^{2j-1}_{3}, \quad j= 2,3,
\cdots
\nonu
\er
where $ t^l_k = w_{l} w_{l-1} \cdots w_{k+1}w_{k}w_{k+1}\cdots w_{l-1}w_{l},
l \geq k, t_k^k = w_k$.  One can prove by induction  its main property, namely
\br
\Pi_0^{(n)}(\a_1) = \a_1, \quad \Pi_0^{(n)}(\a_2) = \a_2 + \a_3 + \cdots +\a_n,
\quad \Pi_0^{(n)}(\a_i) = -\a_{n-i+3}, i=3,4, \cdots n
\nonu
\er
Since $\Pi_0 (E_{\b}^{(k)}) = E_{\Pi_0(\b)}^{(k)}$ we realize that 
\br
\Pi_0 (\eps_+) =m\( E_{-\a_2}^{(1)} + E_{-\a_3}^{(0)} + \cdots 
+E_{-\a_{n-1}}^{(0)} +
E_{-\a_n}^{(0)} +E_{\a_2+\a_3 \cdots +\a_n}^{(0)}\)
\nonu
\er
Remember that $E_{\b}^{(l)}$ can be represented as $E_{\b}^{(l)}= \lambda^{l}
\otimes E_{\b}^{(0)}$ and $\hat d = \l {{d}\o {d\l }}$.  We next define an
operator $T_3 = \l_2 \cdot H^{(0)} + 2\hat d$ such that 
\br
\Omega (E_{-\a_2}^{(1)}) = e^{-ln(\l )T_3}E_{-\a_2}^{(1)}e^{ln(\l )T_3}= {1\o
\l}E_{-\a_2}^{(1)} = E_{-\a_2}^{(0)}
\nonu
\er
\br
\Omega (E_{\a_2 +\cdots +\a_n}^{(0)}) ={1\o \l} E_{\a_2 +\cdots +\a_n}=
 E_{\a_2 +\cdots +\a_n}^{(-1)},\;\;\;\;\;\;\;\;\;\;
 \Omega (E_{\pm \a_k}^{(0)}) =E_{\pm \a_k}^{(0)}, k=3,4, \cdots n
\nonu
\er
Therefore 
\br
e^{-ln(\l )T_3}\Pi_0 (\eps_+)e^{ln(\l )T_3} = \eps_-
\nonu
\er
and  as a result we find that the operator $\Pi$ has the form $\Pi = \Omega \Pi_0 $.

Taking into account the explicit parametrization (\ref{2.6}) and (\ref{5.4}) of $D_a$ we derive
the following field transformations obtained from $D^{\pr} = \Pi (D^{-1})$,
\br
\varphi^{\pr}_k &=& \varphi_{n-k+1} - \varphi_1 - 
{{n-k}\o {\b n}} ln (\Delta_0),\quad 
k=1,2, \cdots , n-1, \quad \varphi_n =0 \nonu \\
R^{\pr} &=& -R + {1\o {\b}}ln {\Delta_0}, \quad \Delta_0 = 
1+ \b^2 \psi \chi e^{-\b \varphi_1} \nonu \\
\psi^{\pr} &=& -\psi e^{-\b \varphi_1} \Delta_0 ^{{{1-n}\o 2n}}, \quad 
\chi^{\pr} = -\chi e^{-\b \varphi_1} \Delta_0 ^{{{1-n}\o 2n}}
\label{2.22}
\er
They leave invariant the potential:$V(D^{\pr}) = V(D)$ and combined with $P$
(i.e. $P\Pi $), generate  symmetries  of the action (\ref{1.1}).
The Lagrangean  (\ref{1.1}) transforms modulo total derivatives ${\cal L}(D^{\pr}) 
= {\cal L}(D) + \pa_{\mu} L^{\mu}$, (see Sect. 8 of ref. \cite{annals}).  
We have to
mention that (\ref{2.22})
 can be obtained from  the vector model  transformations
 (\ref{2.18}) applying the {\it axial-vector} (non-local)
  change of variables  (\ref{2.23}).
As we shall see in Sect. 3, the {\it discrete symmetries }(\ref{2.22}) play a crucial role in
the derivation of the first order `` solitonic '' equations (\ref{3.4}).

{\bf 2.7 Weyl families of IM's.} 
The algebraic (Weyl group) constructions used in the discussion of the discrete
symmetries of (\ref{1.1}) addresses the question of whether the remaining Weyl group
elements ( or their specific combinations including, say $w_{\a_1}$, etc. ) act as
symmetries of our model (\ref{1.1}) and {\it if not}, whether the Lagrangeans
obtained represent {\it new integrable models}.  This problem 
appears to be  the {\it nonconformal }
generalization of the {\it Weyl families } of conformal non-abelian Toda models
constructed in ref. \cite{annals} ( see Sect. 8).  The simplest transformation $D^{\pr} =
w_{\a_1} (D)$ is {\it not a symmetry } of (\ref{1.1}).  It has the following
components form $( D^{\pr} = (\psi_0, \chi_0, \varphi_{0i}))$ 
\br
\varphi_k &=&\varphi_{0k} - {{n-k}\o {\b n}}ln 
( 1+ \b^2 \psi_0 \chi_0 e^{\b \varphi_{01}}) \nonu \\
\psi  &=& \chi_0 e^{\b \varphi_{01}} ( 1+ \b^2 \psi_0 \chi_0 e^{\b \varphi_{01}})^{-{{n-1}\o
{2n}}}\nonu \\
\chi  &=& \psi_0 e^{\b \varphi_{01}} ( 1+ \b^2 \psi_0 \chi_0 e^{\b \varphi_{01}})^{-{{n-1}\o
{2n}}}
\label{2.24}
\er
As a result of this change of variables the $w_{\a_1}$-{\it image } of (\ref{1.1}) is
an integrable model with Lagrangean ${\cal L}_{w_{\a_1}}$,
\br
{\cal L}_{w_{\a_1}} =  {1\o 2} \eta _{ij} \pa \varphi_{0i} \bar \pa
\varphi_{0j}  + {{\pa \chi_0 \bar \pa \psi_0 e^{\b \varphi_{01}}}\o {1+ \b^2{{n+1}\o {2n}} \psi_0 \chi_0
e^{\b \varphi_{01}}}} -  V_0 
\nonu
\er
\br
V_0 = {{m^2}\o {\b^2}} \( e^{\b (-2\varphi_{01} + \varphi_{02})}
 ( 1+ \b^2\chi_0 \psi_0
e^{\b \varphi_{01}}) + e^{\b (\varphi_{01} + \varphi_{0 n-1})} +  
\sum_{k=2}^{n-1} e^{\b (\varphi_{0 k-1}+\varphi_{0 k+1}- 2 \varphi_{0 k})} 
 - n\) 
\nonu
\er
where $\varphi_{0n} = 0$. The other Weyl group elements $w_{\a_i}$ and 
their combinations  lead to transformations similar to
(\ref{2.24}), thus producing new 
families of classically equivalent
integrable
models ${\cal L}_{w_{\a_i}}$.   Hence we  can conclude that $P \Pi
$-transformation given by eqns. (\ref{2.22}) is the unique (affine) Weyl
group transformation leaving (\ref{1.1}) invariant.  Contrary to the abelian
affine Toda theories all the other Weyl transformations (different from $
\Pi $) are not symmetries of (\ref{1.1}), thus giving rise to {\it new 
phenomena}- Weyl families of IM's. 


\sect{Soliton Equations}
{\bf 3.1 Vacua Backlund Transformations.}
Consider two arbitrary solutions $D_1$ and $D_2$ of eqns. (\ref{2.9}).  The
corresponding Lax connections $   {\cal A}_{\mu}^W (s) = 
 {\cal A}_{\mu}^W (D_s) $, $s=1, 2$ are related by appropriate {\it dressing}
(gauge) transformations $\Theta_{\pm}$:
\be
 {{\cal A}_{\mu}^W {(2)} }= \Theta_{\pm}  { {\cal
A}_{\mu}^W {(1)}}{\Theta_{\pm}}^{-1} +
 (\pa_{\mu}\Theta_{\pm} )  {\Theta_{\pm}}^{-1}
\label{3.1}
\ee
 They leave invariant eqn. (\ref{2.2}) together with  the linear
 problem
 \be
 \( \pa_{\mu} - {\cal A}_{\mu}^W (D_s) \) T(D_s) = 0.
 \label{3.2a}
 \ee
   The relation between the monodromy matrices 
 $T_s = T(D_s) = P \exp ( \oint  
{\cal A}_{\mu}^W dx^{\mu} )$ has the well known form \cite{babelon}
\be
T_2 = \Theta_{\pm}T_1, \quad \Theta_{+}T_1 = \Theta_{-}T_1g_0^{(1)}
\label{3.2}
\ee
where $g_0^{(1)} $ is a constant element of the corresponding affine group.
The {\it strategy} in deriving the  infinitesimal Backlund transformations $D_1
\leftarrow \rightarrow D_2$ consists in the following {\bf a)} First solve eqn. 
(\ref{3.2a}) for $\Theta_{\pm}(D_s)$ and {\bf b)} find first  order differential
equations for $D_s$ by substituting these $\Theta_{\pm}(D_s)$ in (\ref{3.1}).
In the case of $A_n^{(1)}$- Abelian affine Toda theories \cite{liao},  and for all
non-abelian Toda theories based on $A_1^{(1)}$, the realization of the above
recipe is quite straightforward.  We find that 
\be
\Theta_+ = X \( 1+ (D_1 X)^{-1}Y D_2 \)
\label{3.3}
\ee
 and the corresponding Backlund transformations  take the following 
  compact form
\br 
D_1^{-1} \pa D_1 X - X D_2^{-1} \pa D_2 &=& [ D_1^{-1} YD_2, \eps_-] \nonu \\
(\bar \pa D_1) D_1^{-1}Y  - Y  (\bar \pa D_2)D_2^{-1} &=& [ D_1X D_2^{-1}, \eps_+]
\label{3.4}
\er
The constant elements $X(\l, a_i), Y(\l, b_i)$ of the universal enveloping
algebra  of $\lie_n^{(a)}$ ($A_n^{(1)}$ or $A_1^{(1)}$ in our case), contain
all the parameters $a_i, b_i$ of the Backlund transformation.  They also have to
satisfy the following conditions 
\be
[X, \eps_-] =0, \quad \quad [Y, \eps_+] =0
\label{3.5}
\ee
It is not difficult to check that the second order equations (\ref{2.9}) are
indeed the {\it integrability conditions} for the first order equations
(\ref{3.4}) if requirements  (\ref{3.5}) are fullfiled.  The verification of
the above statement does not depend on the explicit form  of $\eps_{\pm}$
( and on the parametrization of $D_s$) and therefore is valid for our  dyonic
$A_n^{(1)}$ model (i.e. $\eps_{\pm}$ and $D_s$ given by eqns. (\ref{2.5}) and 
(\ref{2.6})).  It turns out however that the simple form  (\ref{3.3}) of 
 $\Theta_{+}$  takes place only
for specific $g_0^{(1)}$ giving rise to 1-soliton solutions \cite{ggsz2}, i.e. when
$D_1 = e^{i\b_0 \l_1\cdot H}= const $ .  The derivation of the explicit form of the
vacua Backlund transformation ($D_1= $const.) for our model (\ref{1.1})
includes one more complication.  Taking $X$ and $Y$ in the form 
\br
X &=& X_{01} I + nX_{02} \l_1 \cdot H + \sum_{k=1}^{n-1} a_k (\eps_-)^k, 
\nonu \\
Y &=& Y_{01} I + nY_{02} \l_1 \cdot H + \sum_{k=1}^{n-1} b_k (\eps_+)^k    
\nonu
\er
and $D$ as in eqn. 
(\ref{2.6}) we have to further {\it impose } on eqns. (\ref{3.4})
 the requirement of $ \Pi$-{\it symmetry} (\ref{2.22}) in order to get  a
 complete system  of equations.  The result is 
 \br
\pa \varphi_k &=& {{m\g }\o \b}\(e^{\b (\varphi_{k-1}-\varphi_k+ {R\o n}) }
-   e^{\b (\varphi_{n-1}+ {R\o n}) } + {{n-k}\o n} \b^2 \psi \chi e^{-\b {R\o
n}}\) \nonu \\
\bar \pa \varphi_k &=& {{m }\o {\g \b}}\(e^{\b (\varphi_{1}- {R\o n}) }-
e^{\b (\varphi_{k+1}-\varphi_k- {R\o n}) }
 + {{k}\o n} \b^2 \psi \chi e^{-\b {R\o n}}\),  \nonu 
 \er
 $k=1,2, \cdots , n-1, \;
 \varphi_0 = \varphi_n =0$
 \br
 \pa R &=&-m\g \b \psi \chi e^{-\b {R\o n}}, \quad 
 \bar \pa R ={{m \b }\o {\g}} \psi \chi e^{-\b {R\o n}}
 \nonu \\
 \pa \psi &=&-m\g \psi e^{\b (\varphi_{n-1} +{R\o n})}\(1- {{n-1}\o {2n}}\b^2
 \psi \chi e^{-\b (\varphi_{n-1}+ {{2R}\o n})}\) \nonu \\
  \pa \chi &=&-m\g \chi e^{\b (\varphi_{1} -{R\o n})}\Delta,\quad  
 \bar \pa \psi ={m\o {\g}} \psi e^{\b (\varphi_{1} -{R\o n})}\Delta \nonu \\
  \bar \pa \chi &=&{m\o {\g}} \chi e^{\b (\varphi_{n-1} +{R\o n})}\(1- {{n-1}\o {2n}}\b^2
 \psi \chi e^{-\b (\varphi_{n-1}+ {{2R}\o n})}\) 
 \label{3.61a}
\er
where $\g = {{b_2 \l}\o {a_1}}= {{b_1}\o {X_{01} - X_{02}}} = 
{{Y_{01} -Y_{02}}\o {a_2}} = \cdots = e^{-b}$ is the Backlund transformation parameter
and the following chain of {\it algebraic relations } should take place,
\br
e^{-\b \phi_1} - e^{\b \phi_2} &=& e^{-\b \phi_2} - e^{\b \phi_3} = \cdots =
e^{-\b \phi_k} - e^{\b \phi_{k+1}} \nonu \\
 =e^{-\b \phi_{n-1}} - e^{\b \phi_{n}} &=& e^{-\b \phi_{n}}- e^{\b \phi_{1}} -
 \b^2 \tilde \psi \tilde \chi
 \label{3.8}
 \er
We have introduced new variables $\phi_k, \tilde \psi$ and $\tilde \chi$
defined by
\br
\phi_k &=& \varphi_k - \varphi_{k-1} - {R \o n}, k=1, \cdots , n, \varphi_0 =
\varphi_n = 0 \nonu \\
\tilde \psi &=& e^{-{{\b R}\o {2n}}}\psi, \quad 
\tilde \chi = e^{-{{\b R}\o {2n}}}\chi, \quad \phi_1 + \phi_2 + \cdots \phi_n = -R
\label{3.9}
 \er
in order to make evident the paralel with the abelian affine Toda case 
\cite{liao}.
  The {\it algebraic eqns.} (\ref{3.8})
are crucial in the proof of the statement that the second order differential
eqns. (\ref{2.1}) are the {\it integrability } conditions for the first order
system (\ref{3.61a}).  In the case of the $A_{n-1}^{(1)}$ abelian affine Toda
model ($\tilde \psi = \tilde \chi = 0 $ ) the analog of eqns. (\ref{3.8})
appears again as a result of the requirement of the abelian analog of the $\Pi$- symmetry 
(\ref{2.22}), i.e. ($\phi^{\pr}_{k} = - \phi_{n-k+1}$) of the first order
equations.
 They do not play  however the same role as eqns. (\ref{3.8}) in the non-abelian model
(\ref{1.1}), but are indeed essential in the derivation of 1-soliton 
solutions\footnote{ in ref. \cite{liao} they have been used
in the form of first integrals $e^{-\b \phi_k} - e^{\b \phi_{k+1}} = const.$ }
\cite{liao}.

 It is important 
 to mention that, although our first order system (\ref{3.61a})
has rather complicated form 
 ( including the non local field $R$) it can be obtained from the simple
 solitonic  equations of the {\it vector model } (\ref{5.6}) and (\ref{5.7})
applying the integrated form of T-duality transformation (\ref{3.10}).
The same is indeed true for the corresponding 1-soliton solutions.

{\bf 3.2 Soliton Spectrum.} 
The {\it main virtue} of the vacua Backlund transformation  ($D_1 = $ const. )
(\ref{3.4}) is  to provide an elementay proof \cite{liao} of the {\it 
 topological character}  of the soliton {\it energy and momenta} ( and also  of
 the {\it electric charge and the spins} , in our case).  The key point is that
 the derivation of the particle -like {\it soliton spectrum} ($M,  Q_{el},
  Q_{\theta}, Q^{top}_{k}, s $) does not require  the explicit knowledge of
 the 1-soliton solutions.  The conserved charges  depend only 
 on the asymptotics of the fields at $x
 \longrightarrow \pm \infty$ and the specific ``{\it solitonic conservation
 laws}'' \footnote{The precise statement is that the soliton spectrum is
 determined by the boundary conditions (b.c. ) and by the algebra of symmetries
 of the first order (BPS-like) equations (\ref{3.61a}) } 
 encoded in eqns. (\ref{3.61a}).  In order
 to extend the arguments used in the abelian Toda model  \cite{liao} to the
 $A_n^{(1)}$-dyonic model (\ref{1.1}) it is convenient to rewrite eqns. 
 (\ref{3.61a})
 in terms of variables $\phi_p, \tilde \psi$ and $\tilde \chi$ defined in 
 (\ref{3.9})
\br
\pa \phi_p &=& \g {m\o \b}\(e^{-\b \phi_p} - e^{-\b \phi_{p-1}} + \b^2 \tilde
\psi \tilde \chi \d_{p,1}\) \nonu \\
\bar \pa \phi_p &=&  {m\o {\g \b}}\(e^{\b \phi_p} - e^{\b \phi_{p+1}} - \b^2 \tilde
\psi \tilde \chi \d_{p,n}\) , p=1,2, \cdots , n \nonu \\
\pa \tilde \psi  &=& -\g m\tilde \psi e^{-\b \phi_n} \Delta_n, \quad \quad
\pa \tilde \chi  = -\g m\tilde \chi e^{\b \phi_1} \Delta_1 \nonu \\
\bar \pa \tilde \psi  &=&  {m \o {\g}}\tilde \psi e^{\b \phi_1} \Delta_1, \quad
\quad 
\bar \pa \tilde \chi  =  {m \o {\g}}\tilde \chi e^{-\b \phi_n} \Delta_n 
\label{3.11}
\er
where $\Delta_1 = 1 + {1\o 2} \b^2 \tilde \psi \tilde \chi e^{-\b \phi_1}, \;\;  
\Delta_n  = 1 - {1\o 2} \b^2 \tilde \psi \tilde \chi e^{\b \phi_n}$.
We next observe that the first order system (\ref{3.11}) (and (\ref{3.61a}) as
well ) admits the following "solitonic" {\it conservation laws }:
\begin{itemize}
\item {\it non chiral}
\br
&&\bar \pa \( \g e^{-\b \phi_k}\) + \pa \( {1\o {\g}} e^{\b \phi_{k+1}} \) =0,
 k=1, \cdots ,n-1 \nonu \\
&&\bar \pa \( \g e^{-\b \phi_n} \Delta_n \) + \pa \( {1\o {\g}} e^{\b
\phi_{1}}\Delta_1 \) =0
\label{3.12}
\er
\item {\it chiral}
\br
\( \g \bar \pa  + {1\o {\g}} \pa \) e^{\pm \b \phi_p} = 0, \quad p= 1,2, 
\cdots ,n  ,\quad
\( \g \bar \pa  + {1\o {\g}} \pa \) ( \tilde \psi \tilde \chi ) = 0, \quad 
\label{3.13}
\er
\be
\( \g \bar \pa  - {1\o {\g}} \pa \) ln ({{\tilde \psi }\o { \tilde \chi}} )= 0
\label{3.14}
\ee
\end{itemize}
Note that the algebraic relations (\ref{3.8}) have been used in the derivation
of the {\it chiral } conservation currents (\ref{3.13}) and (\ref{3.14}).  The
conclusion is that the {\it complete algebra } of symmetries of the 1-soliton
solutions of the axial model (\ref{1.1}) is generated by the Poincar\'e
currents $T_{\mu \nu}$ (with ``charge ''  the 2-momenta $P_{\mu}$) and $M^{\rho,
\mu \nu } = x^{\mu } T^{\rho \nu } - x^{\nu } T^{\rho \mu }+spin\;\; matrix$
 ( the 2-d spin-orbital
momenta is $i\int ^{\infty}_{-\infty} dx M^{0,01} =s$), the electric  and magnetic currents
$J_{el}^{\mu ,ax},  J_{\theta}^{\mu ,ax}, J_{k}^{\mu}$ given by eqns.(\ref{2.14}),(\ref{jota})
and (\ref{carga})   and by
the {\it ``internal'' currents} (\ref{3.12}), (\ref{3.13}) and (\ref{3.14}). 
Leaving aside  an interesting problem of deriving the explicit form of this
algebra, we will restrict ourselves to consider few simple {\it consequences}
of eqn. (\ref{3.12}) and (\ref{3.13}), that allows us to find the {\it mass spectrum } of
the $U(1)$ {\it charged} 1-solitons.  

Taking into account eqns. (\ref{3.11}) and (\ref{3.12}) we realize that the
potential (\ref{2.8}) 
\br
V&=& {{m^2}\o {\b^2}}\( \sum_{k=1}^{n} e^{\b (\phi_{k+1} - \phi_k)} +
 \b^2 \tilde \psi \tilde \chi e^{-\b \phi_n} -n \)
 \nonu 
 \er
 and the stress-tensor $T_{\mu \nu}$:
 \br
 T_{00} &=& {1\o 4} \sum_{k=1}^{n} \( (\pa \phi_k)^2 + (\bar \pa \phi_k)^2  \)
 + {{e^{-\b \phi_1}}\o {2 \Delta_1}}\( \bar \pa \tilde \psi \bar \pa \tilde \chi  +
  \pa \tilde \psi  \pa \tilde \chi \) + V \nonu \\  
T_{01} &=& {1\o 4} \sum_{k=1}^{n} \( (\pa \phi_k)^2 - (\bar \pa \phi_k)^2  \)
 + {{e^{-\b \phi_1}}\o {2 \Delta_1}}\(  \pa \tilde \psi  \pa \tilde \chi 
  -\bar \pa \tilde \psi  \bar \pa \tilde \chi \)   
  \er
  are {\it total derivatives}:
  \br
  V  = -{1\o 2}\( \bar \pa F^- + \pa F^+ \), \quad
  T_{00}  = \pa _{x } \(  F^- -  F^+ \), \quad  
  T_{01}  = \pa _{x } \(  F^- +  F^+ \)
  \label{3.15}
  \er
  The $F^{\pm}$ we have  introduced in eqn. (\ref{3.15}) turn out to be the
  light cone components $F_{\mu} = (F^-, F^+ )$ of certain linear combination of the conserved
  currents (\ref{3.12}):
\br
F^- = -{{m \g }\o {\b^2}} \sum_{k=1}^{n} e^{-\b \phi_k}, \quad F^+ = {{m}\o
{\g \b^2}} \( \sum_{k=1}^{n} e^{\b \phi_k} + \b^2 \tilde \psi \tilde \chi \)
\label{3.16}
\er
Hence the {\it energy} and the {\it momentum}  of the 1-soliton decribed by
eqns. (\ref{3.11}) receive contributions from the {\it boundary terms } only:
\be
E= \int _{-\infty}^{\infty} T_{00} dx = \(F^- - F^+ \)|_{-\infty}^{\infty},
\quad 
P= \int _{-\infty}^{\infty} T_{01} dx = \(F^- + F^+ \)|_{-\infty}^{\infty}
\label{3.17}
\ee
The same is true for the {\it electric} charge (\ref{2.14}) ,(\ref{cargael})
\br
Q_{el}^{ax} = \int _{-\infty}^{\infty}J_{el}^{0, ax} dx = 2\b_0 \( R(\infty ) - R(-\infty ) \)
\nonu 
\er
\br
\int _{-\infty}^{\infty}J_{el, impr}^{0, ax} dx = Q_{el}^{ax} - {{\b_0^3}\o {4\pi^2}}\sum_{ k=1}^{n-1}
\nu_k (\varphi_k(\infty ) -\varphi_k(-\infty ))
\label{cgel} 
\er
as well as for the {\it topological currents } $  J^{\mu ,ax}_{\theta} $
and $J^{\mu}_k $ by construction.
 We fix the asymptotics of the fields $\phi_p, \tilde \chi ,  \tilde  \psi$ ( and $R$) at $x
 \longrightarrow \pm \infty $ by requiring $V|_{x\rightarrow \pm \infty} =
 0$, i.e. the solutions of (\ref{3.11})  we seek for to interpolate between
 two nontrivial vacua  (\ref{2.12}).  More precisely, we choose,
 \br
 \varphi_l (\pm \infty) = {{2 \pi}\o \b_0} {l\o n} N_{\pm}, \;\; l=1,2, \cdots
 ,n-1, \quad 
  \tilde  \chi  \tilde  \psi (\pm \infty) = 0, \quad R(\pm \infty)= {{2\pi}\o {\b_0}}f_{\pm}
 \nonu 
 \er
 and \footnote{The precise definition \cite{ggsz2} of the topological charge $Q_{\theta}$ for arbitrary
 complex $\psi$ and $\chi$ is $Q_{\theta} = Re \( \theta (\infty ) -
 \theta (-\infty )\), \;\; Re \theta (\pm \infty ) = {1\o {2\b_0}} \( Arg \tilde \chi - 
 Arg \tilde \psi \)|_{\pm \infty}$.For the 1-soliton solutions it coincides with
 (\ref{3.18})}
 \be
   \theta (\pm \infty ) = -{{i\o {2\b_0 }}}ln({{\tilde  \chi} \o {\tilde  \psi}})|_{\pm
  \infty} = {{\pi }\o {2\b_0} }L_{\pm} 
  \label{3.18}
  \ee
  where $N_{\pm} $  and $L_{\pm}$ are arbitrary integers and $f_{\pm}$  are real numbers.
According to eqn. (\ref{3.9}) we also get 
\be 
\phi_p (\pm \infty) = {{2\pi} \o {n\b_0}} ( N_{\pm} - f_{\pm} ), 
\label{ass}
\ee
 Substituting (\ref{ass}) into (\ref{3.17}) for
$\b = i\b_0 $ we derive the 1-soliton {\it energy momentum
spectrum} \footnote{ The complex form of $E$ and $P$ is misleading.  It will be shown below 
that the chain relations
(\ref{3.8}) together with the conservation laws (\ref{3.12}), (\ref{3.13}) and (\ref{3.14})
impose conditions on $N_{\pm}$ and $f_{\pm}$ that ensures the reality of $E$ and $P$, ($E \geq 0$,
as well )} :
\br
E = {{4m}\o {\b_0^2}}n \sin {{1}\o {4n}} \( Q_{el}^{ax} -\b_0^2 Q_{mag}\)\sin a \nonu \\
P = {{4mi}\o {\b_0^2}}n \sin {{1}\o {4n}} \(  Q_{el}^{ax} -\b_0^2 Q_{mag}\)\cos a 
\label{3.19}
\er
where $a= {{\pi}\o n} ( N_+ + N_- - f_+ -f_- ) - ib$, $ \g = e^{-b}$.  The
{\it soliton charges } are given by:
\br
Q_{el}^{ax} &=& 4\pi \(f_+ - f_- \), \quad 
j_{\varphi} =  N_+ - N_- =0,\pm 1,\pm2, \cdots ,\pm(n-1) ,\quad
Q_{mag} = {{4\pi}\o {\b_0^2}} j_{\varphi}
\nonu
\er
It turns out that the masses of the charged 1-solitons of the {\it axial model}
(\ref{1.1}):
\be
M^{ax} = \sqrt{E^2 - P^2}= {{4m}\o {\b_0^2}}n |\sin {{1}\o {4n}}(Q_{el}^{ax} - 
\b_0^2 Q_{mag} )|
\label{3.21}
\ee
does not depend on the topological charge $ Q_{\theta}^{ ax}$. 
 As we shall show in Sect.5 the same formula 
(\ref{3.21}) takes place for the 1-solitons of the {\it vector model } 
(\ref{1.3}) but with $Q_{el}^{ ax}$ replaced by its dual  
$ Q_{\tilde \theta}^{vec} = -{i}\int_{-\infty
}^{\infty} \pa_{x} ln (A^2)$ and $ Q_{el}^{vec} = -{i}\int_{-\infty
}^{\infty} \pa_{x}ln (\tilde u^2)=0$.  
Note that   the 1-soliton
spectrum of the dyonic model (\ref{1.1}) (i.e., $M, Q_{el},  Q_{\theta},
Q_k, s, E $)  depends upon  arbitrary parameters $f_{\pm}$ (i.e.
on  the b.c. of the nonlocal field $R$).  They are not related to the b.c. of the
{\it physical } fields
$\varphi_l, \psi, \chi $ and it does not seems to be one of the new {\it
internal } conserved charges.  We are going to show now that $f_{\pm}$
represent the value of the following chain of {\it first integrals } of eqn. 
(\ref{3.11})
\be
y= e^{-i\b_0 \phi_p } - e^{i\b_0 \phi_{p+1} } = e^{-i\b_0 \phi_n } - e^{i\b_0 \phi_{1} }
+ \b_0^2 \tilde \psi \tilde \chi, \quad p=1,\cdots ,n-1
\label{3.23}
\ee
It is easy to check that as a consequence of eqns.  (\ref{3.8}), (\ref{3.12})
and (\ref{3.13}) we have 
\be
\bar \pa \( e^{-i\b_0 \phi_1} - e^{i\b_0 \phi_2} \)=0, \quad \bar \pa \(
 e^{-i\b_0 \phi_n}- e^{i\b_0 \phi_1} + \b_0^2 \tilde \psi \tilde \chi \)=0, \quad etc
\nonu 
\ee
Combining them with the chiral conservation laws (\ref{3.12})
and (\ref{3.13}) we realize that all the
members of the chain relations (\ref{3.8}) represent first integrals
of the system (\ref{3.23}).  Thus, the only (``zero'') mode of the {\it double
chiral } (i.e., $\pa y = \bar \pa y =0$) conserved currents $y = 2i \sin \a $
 describes an internal conserved
charge.  Since (\ref{3.23}) is valid for all $x$ and $t$ applying them to the
boundary case ($t$ fixed and $x \rightarrow \pm \infty $) we find  the
following relations between $f_{\pm}, N_{\pm}$ and $\a $:
\be
\sin{{2\pi }\o n}(f_{\pm}- N_{\pm})= \sin \a 
\label{3.24}
\ee
  The general solution of (\ref{3.24}) is given by 
\be
{{2\pi }\o n}(f_{\pm}- N_{\pm})= \a + 2\pi S_{\pm}^{(a)},  
\nonu
\ee
or 
\be
{{2\pi }\o n}(f_{\pm}- N_{\pm})= Sign (\a ) \pi -\a + 2\pi S_{\pm}^{(b)},  
\nonu
\ee
$S_{\pm}^{(a,b)}= 0, \pm 1, \pm 2, \cdots $.
We chose $f_{\pm}$ (for $cos \a > 0 $) ,such that they provide  a nontrivial $\a$-dependence of
$Q_{el}$:
\br
f_{+} &=& {n \o {2\pi }}(\a + 2\pi S_+^{(a)}) + N_+ \nonu \\
f_{-} &=& -{n \o {2\pi }}(\a -\pi Sign (\a )+ 2\pi S_-^{(b)}) + N_-
\label{3.25}
\er
(and $f_+ \leftrightarrow f_- ,\quad N_+ \leftrightarrow N_- $ for $ cos \a < 0$ )
and therefore
\br
&&Q_{el}^{ax} = 4\pi  (f_+ -f_-)  = 4n (\a -
{\pi \o 2}Sign (\a )) + 4\pi  j_{\varphi }, 
\nonu \\
&&\int_{\infty}^{\infty} J^{0, ax}_{el, impr} dx =
 Q_{el}^{ax} - {{\nu  \b_0^2}\o {2\pi}} j_{\varphi}, 
\quad \quad 
\label{3.26}
\er
neglecting the term ${n} (S_+^{(a)} +S_-^{(b)})$ in $Q_{el}^{ax}$ since $j_{\varphi
}$ is defined as integer mod $n$. 

Taking into account (\ref{3.25}) we realize that 
\br
\sin a = -Sign ( \a ) \cosh b, \quad \cos a = -i Sign (\a )\sinh b
\nonu
\er
and therefore
\br
E= M^{ax}\cosh b, \quad P=-M^{ax}\sinh b, \quad M^{ax}={{4mn}\o {\b_0^2}}|\cos \a |
\nonu 
\er
(i.e., $E\geq 0$ as it should be ).  To make the discussion of the spectrum of
the $U(1)$ -charged topological soliton complete we antecipate the
semiclassical quantization of $Q_{el}$ (see Sect. 7.1 below):
\be
Q_{el}^{ax} = \b_0^2(j_{el}+ {{\nu}\o {2\pi}}j_{\varphi}), \quad j_{el}= 0, \pm 1,
\cdots
\label{qeletrica}
\ee
This form of $Q_{el}^{ax}$ confirms the arguments presented in Sect. 2 that the
electric charge of the solitons (and breathers as well \cite{ggsz2})  of the
axionic model (\ref{1.1}) gets contributions from the magnetic charge $Q_{mag}
= {{4\pi}\o {\b_0^2}} j_{\varphi}$.  It is important to note that the
1-soliton charges ($Q_{el}^{ax}, Q_{mag}$) (\ref{qeletrica}) coincide with the
electric and magnetic charges of the dyonic solutions of 4-D \\
$SU(n+1)$ YMH model
with CP -breaking $\theta $-term \cite{witten1}.  The 2-D soliton mass spectrum
(\ref{1.5})  however is different from the semiclassical masses $M^2_{dyon} =
 <\phi >(Q_{el}^2 + Q_{mag}^2)$ of 4-D YMH dyons.  It is worthwhile to mention 
 that the $n \rightarrow \infty$ limit of our mass formula
 (\ref{3.21}) coincides with the BPS bounds for the masses of particular dyons 
 ($Q_{el}^{(2)} = 0 = Q_{mag}^{(1)}$) of four dimensional
 $N=2$ Supersymmetric YM theory \cite{gauntlett}.

\sect{Electrically Charged Topological Solitons}
{\bf 4.1 Soliton Solutions.}
In our derivation of the {\it dyonic properties}  (\ref{1.5}) of the 1-solitons
of (\ref{1.1}) we left unanswered the {\it important question }: whether eqs. 
 (\ref{2.1}) and (\ref{3.11}) possess soliton solutions with both charges
 $Q_{el}$ and $Q_{m}$ {\it different from zero}, i.e. $N_+ \neq N_-$ and $f_+
 \neq f_-$.  It is indeed the case as we will show by explicit construction of
 solutions of eqns. (\ref{3.11}) and (\ref{3.8}).  We first consider the
 $\phi_p, p= 1, \cdots ,n $ equations.  Taking into account the algebraic
 relations (\ref{3.8}) and (\ref{3.23}) we realize that they can be rewritten in
 the following compact form:
 \br
 \pa \( e^{i\b_0 \phi_p} \) &=& me^{-b} \(1 - e^{2i\b_0 \phi_p} - 
 2i\sin\a e^{i\b_0 \phi_p}
 \) \nonu \\
 \bar \pa \( e^{i\b_0 \phi_p} \) &=& -me^{b} \(1 - e^{2i\b_0 \phi_p} - 
 2i\sin\a e^{i\b_0 \phi_p}
 \)
 \nonu
 \er
 We next introduce the ``solitonic'' variables
 \br
 \rho_+ &=& \cosh (b)x - \sinh (b)t, \quad \pa_{\rho_{+}} = \sinh (b)\pa_{t} +
 \cosh (b) \pa_{x} \nonu \\
 \rho_- &=& \cosh (b)t - \sinh (b)x, \quad \pa_{\rho_{-}} = \cosh (b)\pa_{t} +
 \sinh (b) \pa_{x}\nonu
 \er
 and taking into account the chiral conservation laws (\ref{3.13})-(\ref{3.14})
 we get 
 \br
 \pa_{\rho_{+}} \( e^{i\b_0 \phi_p} \) = m \(1 - e^{2i\b_0 \phi_p} - 
 2i\sin(\a)e^{i\b_0 \phi_p} \),\quad \quad
 \pa_{\rho_{-}} \( e^{i\b_0 \phi_p} \) =0
 \label{4.2}
 \er
 The general solution of eqns. (\ref{4.2}) is given by
 \be
 e^{i\b_0 \phi_p} = e^{i\a } {{S_p e^{-2i\a +2m\cos \a \rho_+} -1}\o 
 {S_p e^{2m\cos \a \rho_+} +1}}
 \label{4.3}
 \ee
 where $S_p$ are certain integration constants satisfying the recursive 
 relations
 \br
 S_{p+1} = e^{-2i\a \pm i\pi } S_p
 \nonu
 \er
 as a consequence of eqns. (\ref{3.23}) and (\ref{4.3}).  Therefore
 \br
 S_p = (-1)^{p-1} e^{-2i\a (p-1)} S_1 \equiv (-1)^{p-n}e^{-2i\a (p-n)}\d
 \nonu
 \er
 and eqn. (\ref{4.3}) acquires the final form
 \be
 e^{i\b_0 \phi_p} = e^{i\a }{{(-1)^{p-1}S_1 e^{-2i\a p} e^{2m \cos \a \rho_+}
 -1}\o {(-1)^{p-1}S_1 e^{-2i\a (p-1)} e^{2m \cos \a \rho_+} +1}}
 \label{4.4}
 \ee
 With eqn. (\ref{4.4}) at hand we can write the 1-solitons in the original
 variables $\varphi_l$ and $R$,
 \be
 R = -\sum_{p=1}^{n}\phi_p, \quad \varphi_l = {l \o n}R + \sum_{p=1}^l \phi_p 
 \nonu
 \ee as follows
 \be
 e^{i\b_0 R} = e^{-in (\a -\pi Sign (\a ) )}{{e^{-f} + 
 (-1)^{n} e^{2in \a}e^f}\o {e^{-f} +
 e^f}}
 \label{4.5}
 \ee
 \be
 e^{i\b_0 \varphi_l} = {{e^{-f} + (-1)^{l-n} e^{2i \a (n-l)} e^f} \o {{(e^{f} +
 e^{-f})}^{{l \o n}}\( e^{-f} + (-1)^{n} e^{2i \a n}e^f \)^{{n-l}\o n}}},
 \quad l=1,
 \cdots n-1
 \label{4.6}
 \ee
 where $\d $ is a complex constant, $f=m\rho_+\cos \a  + {1\o 2} ln (\d )$. 
  The next step is to
  derive solutions for $\psi$ and $\chi $.
   Eqns. (\ref{4.5}) and (\ref{4.6}) together with the algebraic relation 
(\ref{3.23}) allows us to determine the product $\psi \chi$:
\br
\b^2 \psi \chi &=& N^2 (e^f + e^{-f})^{-{{n+1}\o n}} \( e^{-f} +
 (-1)^n e^{2i\a n
}e^f \)^{-{{n-1}\o n}} \nonu \\
 N^2 &=& (1 + e^{-2i\a})((-1)^n e^{2i\a n} -1)
 \label{4.7}
 \er
 It turns out that the ratio ${{\tilde \psi} \o {\tilde \chi}} = 
 {{ \psi} \o { \chi}}$ satisfies the following simple equations
 \br
 \pa ln ({{ \psi} \o { \chi}}) = me^{-b} \(e^{i\b_0 \phi_2} - e^{-i\b_0 \phi_1} \),
 \quad
 \bar \pa ln ({{ \psi} \o { \chi}}) = me^b \(e^{i\b_0 \phi_2} - e^{-i\b_0 \phi_1} \)
 \nonu
 \er
 Applying once more eqn. (\ref{3.23}) we find that $ln ({{ \psi} \o { \chi}})$
 is independent of $\rho_+$ 
 while the $\rho_-$dependence  is linear
 \be
 \pa_{\rho_-} ln ({{ \psi} \o { \chi}}) =  -2im \sin \a
 \nonu 
 \ee
 Therefore the ratio $ {{ \psi} \o { \chi}}$ is given by
 \be
 {{ \psi} \o { \chi}}= e^{-2i(m\sin \a \rho_- + q)}
 \label{4.9}
 \ee
where $q$ is an arbitrary complex constant.  
Thus, eqns.  (\ref{4.7}) and (\ref{4.9})
 completely determine the solution of the 
 first order equations  (\ref{3.61a})
 \br
 \psi &=& \pm {{N\o \b}}e^{-i (m\sin \a \rho_- +q)} (e^f + e^{-f})^{-{{n+1}\o
 {2n}}} \( e^{-f} + (-1)^n e^{2i \a n } e^f \) ^{-{{n-1}\o {2n}}}
 \nonu \\
 \chi &=& \pm {{N\o \b}}e^{i (m\sin \a \rho_- +q)} (e^f + e^{-f})^{-{{n+1}\o
 {2n}}} \( e^{-f} + (-1)^n e^{2i \a n } e^f \) ^{-{{n-1}\o {2n}}}
 \label{4.10}
 \er
 The 1-{\it soliton solution } of the {\it dyonic } model (\ref{1.1}) is
 presented by the set of functions (of $\rho_+$ and $\rho_-$)  (\ref{4.6})
 and (\ref{4.10}).

 {\bf4.2 Soliton Charges.}
 It remains to be shown that these solitons carry non trivial
 electric and magnetic charges.  The simplest way to do this consists in
 calculating the asymptotics of the fields $R, \varphi_l, \psi , \chi $ from
 eqns. (\ref{4.5}), (\ref{4.6}) and (\ref{4.10}) and  further comparing with the values
  (\ref{3.18}) and \ref{3.25}) proposed in Sect. 3.
 We have to distinguish three cases \footnote{ We are not considering here two
 limiting cases: a) $\cos \a =0$ since  there are no true 1-soliton and b)
 $\sin \a =0$, $n$ even which leads to $N^2 =0$ and $\psi =\chi =0$ (i.e. to
 the $A_n^{(1)}$ abelian Toda model) }: i)$\cos \a >0$; ii))$\cos \a <0$; iii)
 $\sin \a =0$, $n$ odd.  Taking the limit $x \longrightarrow \pm \infty $ in
 eqns.  (\ref{4.5}), (\ref{4.6}) and (\ref{4.10}) we obtain  for $\cos \a >0$:
 \br
 \varphi_l (\pm \infty ) &=& {{2\pi }\o \b_0 } \({l\o n} N_{\pm }^l + K_{\pm
 }^l\)\quad ,\theta (-\infty ) = \theta (\infty ), \quad j_{\theta} =0
 \nonu \\
 R ( \infty ) &=&{{2\pi }\o \b_0 }f_+ = {{2\pi }\o \b_0 } \( {{n\a } \o {2\pi }}
 + S_+\), \nonu \\
  R ( -\infty ) &=&{{2\pi }\o \b_0 }f_- = {{2\pi }\o \b_0 } \( 
  -{{n(\a -\pi Sign (\a) )} \o {2\pi }}
 + S_- \),
 \label{4.11}
 \er 
 The arbitrary integers $N_{\pm}^l,K_{\pm}^l $ and $S_{\pm}$ are further
 restricted by the conditions 
 \begin{enumerate}
 \item to provide nontrivial zeros of the potential (\ref{2.8})
 \item chain relations (\ref{3.8}) and (\ref{3.23})
 \end{enumerate}
 The simplest solution  satisfying 1) and 2) that leads to nontrivial {\it
 dyonic spectra} (\ref{3.26})(with $j_{\theta }=0$) is given by 
 \br
 N_{\pm}^l = N_{\pm} (mod \;\; n), \quad \quad S_{\pm} = N_{\pm} (mod \;\; n), \quad \quad
 K_{\pm}^l =0
 \nonu
 \er
 for all $l=1, 2, \cdots, n-1$ ($N_{ \pm}$ being new arbitrary integers). 
  The important fact is that the 1-soliton solution given by 
 (\ref{4.5}), (\ref{4.6}) and (\ref{4.10}) (for $\cos \a >0$) are topological
 and electrically charged, i.e., with both charges $Q_{el}^{ax}$ and $Q_m$ different
 from zero.  Such {\it dyonic type soliton} combines   the
 properties of the Lund-Regge (complex SG) solitons \cite{dorey} with the 
 $A_n$-abelian affine Toda   solitons \cite{olive1},\cite{liao}.  
  The case $\cos \a <0$ leads
 again to (\ref{4.11}) but with $N_+^l \leftrightarrow N_-^l$ and $f_+ 
 \leftrightarrow f_-$, i.e. $Q_{el} \rightarrow -Q_{el}$ and 
 $Q_{mag} \rightarrow -Q_{mag}$. 
  Therefore  the corresponding "particles " can be interpreted as antisolitons. 
  
  {\bf 4.3 Periodic lumps.}
  The {\it
 main feature} of the 1-solitons with $\sin \a \neq 0$ 
 (and $\cos \a \neq 0$) is that in the rest frame $v= \tanh b=0$, they
 represent {\it periodic } in time $t$,  ($\tau = {{2\pi} \o {\omega}},
 \omega = m\sin \a$) bounded solutions.  Such ``periodic lumps'' behaviour is
 familiar from the CSG solitons \cite{dorey} and the breathers ( doublet solutions
 ) of the SG theory \cite{dash}.  It reflects the angular property of the phase 
 $2i\b_0 \theta  = ln ( {{ \tilde \chi} \o { \tilde \psi}})$
 of the 
 fields $\psi, \chi $,   (see
 eqn. (\ref{4.9})):
 \be
 \theta (t +\tau , x ) = \theta (t, x) -{{ 2 \pi} \o \b_0}
 \label{4.13}
 \ee
 and therefore $t \in S_1$ and $\theta \in S_1$.  The particle interpretation
 of the 1-{\it soliton data}: center of mass $X = Re {1\o 2} ln (\d )$, $U(1)$-
 moduli $q$, its velocity $v = \tanh (b)$ and the angular velocity $\omega = m
 \sin \a$ can be borrowed from the CSG model \cite{dorey} as particle with internal
 coordinate $q$, rotating in the $q$ space with constant angular velocity
 $\omega $ at the rest frame $v =0$.  Another interpretation comes from the SG
 breather, as bounded motion of charged particles \cite{dash}.  The singular case
 $\sin \a =0$, i.e. $\a = s \pi,\;\;  s=0,\pm 1,\pm 2, \cdots $ and 
 $n$-odd (i.e. $N^2 \neq 0$) 
is an example of static solitons ($v=0$) with charges
\br
Q_{mag} = {{4\pi}\o {\b_0^2}} j_{\varphi}, \quad \quad Q_{el}^{ax}= 4n \pi \((s- {1\o
2})sign (cos\a) + 
 {{j_{\varphi}\o n}}\)
\nonu
\er
and degenerate (independent of charges) mass
$ M= {{4m}\o \b_0^2}n $.

 {\bf 4.4 Breathers.} 
 It is important to note that the electrically charged topological 1-soliton
 solution of our model {\it do not exhaust} all the {\it particle-like }
 solutions of (\ref{1.1}).  The {\it complete list} of (topologically ) stable 
 1-solitons, breathers and breathing ( or ``excited solitons '' ) includes together
 with the above constructed $U(1)$-{\it charged topological } 1-soliton (2-D
 dyon) also the {\it neutral} 1-soliton of specie  $d$, $d=1, \cdots ,n-1$
  ($M_d, Q_{mag} =
 {{4\pi}\o {\b_0^2}}d , Q_{el}^{ax}
 =0, Q_{\theta}^{ax} =0$, i.e. monopole ) and their bound states:
 \begin{itemize}
 \item The $A_{n-1}^{(1)}$ abelian affine Toda breather (two neutral vertices
 \cite{olive1}) and the ``excited solitons ''. 
 \item The (NA Toda ) three vertex breathing \cite{ggsz2} describing the one charged
 1-soliton and one neutral ($d$) 1-soliton bound state.
 \item The four vertex charged breathers \cite{ggsz2} describing the bound state of
 two charged topological 1-solitons.
 \end{itemize}
 The explicit construction of all these solutions for the axial model
 (\ref{1.1}) as well as their spectrum are presented in our forthcoming
 publication \cite{ggsz2}.  The method employed in \cite{ggsz2} is a slight modification
 of the standard (abelian Toda ) ``vertex operators ''  \cite{babelon} (or soliton
 specialization \cite{olive1} or Hirota $\tau$ function\cite{aratyn}) method adapted to the
 case of the {\it singular } Non-Abelian affine Toda models (\ref{1.1}) and 
 (\ref{1.3}). 
  The 3-vertex breather represent an interesting example of $U(1)$
 charged particle-like solution carrying both topological charges $j_{\varphi}$
 and $j_{\theta}=\pm 1$ (remember  that $j_{\theta}=0$ for our charged solitons 
 (\ref{4.5}), (\ref{4.6}),(\ref{4.10}) ) \cite{ggsz2}:  
 \br
 Q_{el}^{ax} &=& \b_0^2 \( j_{el} + {{\nu \o {2\pi }}}(j_{\varphi} + d)\), 
 \quad Q_{mag} = {{4\pi \o {\b_0^2}}}(j_{\varphi} +d) \nonu \\
 Q_{\theta} &=& {{\pi }\o {2}}\( j_{\theta} + 
 4 \tilde \g j_{\varphi}\) , \quad j_{\theta} = \pm 1\quad; 
 j_{\varphi}, d = \pm 1, \cdots ,\pm (n-1).
 \label{4.15}
 \er
   Due to  the fact that
 $Q_{\theta}^{ax} \neq 0$ is the T-dual of the vector model electric charge $ Q_{el}^{vec}\neq 0$
 ($Q_{el}^{ax}$ is the dual  of the vector model topological charge 
 $ Q_{\tilde \theta}^{vec}= -2i\int dx \pa_{x} ln (A)$),  this solution plays an important role in the
 understanding of the T-duality relations between the discrete spectra of
  the axial and vecor IM's.

\sect{Vector Model Solitons}

 {\bf 5.1 First order Equations.}
 Taking $D_1 = D_v, D_2 = e^{\b \l_1 \cdot H}$ and $X = X_{01} I + nX_{02} \l_1 \cdot H +
 \sum_{k=1}^{n-1} a_k (\eps_-)^k, \;\; 
 Y = Y_{01} I + nY_{02} \l_1 \cdot H +
 \sum_{k=1}^{n-1} b_k (\eps_+)^k$ in eqn. (\ref{3.4}) we derive the
 following incomplete system of first order equations 
 \br
 \pa B &=& m\g (1-AB) c_1 c_2 \cdots c_{n-1}, \quad \quad 
 \bar \pa A = -{{m}\o {\g}} {{(1-AB)}\o { c_1 c_2 \cdots c_{n-1}}}
  \nonu \\
 \pa ln (c_1 ) &=& m \g \( {1 \o {c_1}} - A c_1 c_2 \cdots c_{n-1}\), \quad
 \quad \bar \pa ln (c_1 )= {{m}\o {\g}}(c_1 - c_2) \nonu \\
 \pa ln (c_k ) &=& m\g \({1 \o {c_k}} - {1 \o {c_{k-1}}}\), \quad
 \quad \bar \pa ln (c_k )= {{m}\o {\g}}\( c_k - c_{k+1}\)
  \nonu \\
  \pa ln (c_{n-1} )&=& m\g \({1 \o {c_{n-1}}} - {1 \o {c_{n-2}}}\)\quad
 \quad \bar \pa ln (c_{n-1} )= {{m}\o {\g}}\( c_{n-1} - {{B}\o {c_1 c_2 \cdots
 c_{n-2}}}\)
 \label{5.6}
 \er
 $k= 2, \cdots ,n-2$, where $\g = {{b_2 \l}\o {a_1}} = 
 {{b_1}\o {(X_{01} - X_{02})}}= \cdots
 =e^{-b}$ is an arbitrary parameter related to the soliton velocity $v$.
 We next impose the condition of invariance of (\ref{5.6}) under the discrete
 symmetries (\ref{2.18}).  It requires that the equations 
 \be
 \bar \pa B = -{m\o {\g}}\(1-AB\) c_1 c_2 \cdots c_{n-1}, \quad \quad \pa A =
 m\g {{(1-AB)}\o {c_1 \cdots c_{n-1}}}
 \label{5.7}
 \ee
 and the following  chain of algebraic relations 
 \be
 Ac_1 c_2 \cdots c_{n-1}- c_1 = {1\o {c_1}} -c_2 = \cdots  = {1\o
 {c_{n-2}}}-c_{n-1} = {1\o {c_{n-1}}} - {{B}\o {c_1 c_2 \cdots c_{n-1}}}
 \label{5.8}
 \ee
 to be satisfied.  An easy check confirms that each set of $A,B, c_i$ that
  solves (\ref{5.6}) and (\ref{5.7}) and 
  \be
  {{B}\o {c_1 c_2 \cdots c_{n-1}}} + Ac_1 c_2 \cdots c_{n-1} = {1\o {c_{n-1}}}
  +c_1
  \label{5.9}
  \ee
  satisfies the second order equations (\ref{5.1}) as well.  The remaining
  algebraic relations ensure the compatibility of (\ref{5.9}) and (\ref{5.6}),
   (\ref{5.7}), i.e. they can be obtained by differentiating (\ref{5.9}) and
   then simplifying with help of (\ref{5.6}) and (\ref{5.7}).
   
    {\bf 5.2 Solitons.}
 An important property of the first order system (\ref{5.6}) and (\ref{5.7}) is
 the existence of specific  ``solitonic '' conservation laws:
 \begin{itemize}
 \item {\it nonchiral }
 \br
 \bar \pa \( \g Ac_1 c_2 \cdots c_{n-1}\) + \pa \( {1\o \g } c_1 \)&=&0, 
 \quad
 \bar \pa \( {{\g }\o {c_{n-1}}} \) + 
 \pa \( {1\o \g }{{B}\o {c_1 c_2 \cdots c_{n-1}}}\)
  =0\nonu \\
 \bar \pa \( \g c_k^{-1}\) + \pa \({1\o \g} c_{k+1}\) &=&0, \quad k=1,2, 
 \cdots, n-1 
 \label{5.10} 
 \er
 \item {\it chiral }
 \br
 && \( \g \bar \pa + {1\o \g } \pa  \) d =0, \quad \quad d=A,B, c_k \nonu \\
 && \( \g \bar \pa - {1\o \g } \pa  \) {\tilde u} =0, \quad \quad 
 \tilde u = {{u \( c_1 c_2 \cdots c_{n-1}\) ^{1\o 2}}\o {\( AB-1 \)^{1\o 2}}}
 \label{5.11}
 \er
 \end{itemize}
 As a consequence of (\ref{5.10}), (\ref{5.11}) and (\ref{5.8}) we find that 
 \br
 \bar \pa \({1\o {c_1}} - c_2\) =0, \quad \quad 
 \bar \pa \( Ac_1 c_2 \cdots c_{n-1} - c_1 \) =0, etc. 
 \nonu 
 \er
  and therefore all the members of the chain (\ref{5.8}) represent first
  integrals of the system (\ref{5.6}) and (\ref{5.7}):
  \br
  {1\o {c_k}} - c_{k+1} = y, \quad Ac_1 c_2 \cdots c_{n-1}- c_1 =y ,\quad \quad
  {1\o {c_{n-1}}}- {{B}\o {c_1 c_2 \cdots c_{n-1}}} =y
  \label{5.12}
  \er
  where $y=2i \sin \a$ is an arbitrary constant.   We next substitute  
 (\ref{5.11}) and (\ref{5.12}) in eqns. (\ref{5.6}) and (\ref{5.7}) resulting
 the following set of $2(n+2)$ first order differential equations for $A,B, c_k,
 \tilde u$:
 \br
 \pa _{\rho_+} c_k &=& 2m\g \(1- c_k^2 - b c_k \), \quad \pa _{\rho_-}c_k
 =0\nonu \\
  \pa _{\rho_+} A &=&{{2m\g }\o {\a_0^{(n-1)}}} \(1- (-1)^{n-1} A^2 - A
  (\a_0^{(n-2)} +\a_0^{(n)}) \),
  \quad \pa _{\rho_-}A =0\nonu \\
 \pa _{\rho_+} {\tilde u} &=&0, \quad  \quad \pa _{\rho_-}{\tilde u} =
  -{{my}\o 2}{\tilde u}
\nonu
\er
where $\a_0^{(s)} = -i {{\sin [(s+1)(\a + {\pi \o 2})}]\o {\cos \a}} $.  
The equation for $B$  is similar  to the one for  $A$.   The following
algebraic relation 
\br
B= {{(-1)^{n-1} A + \a_0^{(n)} }\o {1- \a_0^{(n-2)}A}}
\nonu 
\er
holds.
In fact, one has to solve equations for $c_k$ only since $A$ and $B$ can be
obtained from (\ref{5.8}).  Applying once more the methods used for the axial
model (\ref{1.1}) in Sect. 4, we get the 1-{\it soliton solutions } for the
vector model  (\ref{1.3}):
\br
c_k &=& e^{i(\a -\pi Sign (\a ) )} {{e^{-f} - (-1)^{k-n} e^{2i \a (n-k-1)} e^f }\o {
 e^{-f} + (-1)^{k-n} e^{2i \a (n-k)} e^f }}, \nonu \\
 A &=& e^{-in(\a -\pi Sign (\a ) )}{{e^{-f} + (-1)^{n} e^{2i \a n} e^f }\o {
 e^{-f} +  e^f }}, \nonu \\
 B &=& e^{in(\a -\pi Sign (\a ) )}{{e^{-f} -  e^{-2i \a } e^f }\o {
 e^{-f} +  (-1)^{1-n} e^{2i \a (n-1)}e^f }}, \nonu \\
 \tilde u  &=& e^{-i (m \rho_- \sin \a + q)}
 \label{5.14}
 \er
 where $f = \rho_+ m \cos \a  + {1\o 2} ln (\d )$ 
 
  {\bf 5.3 Soliton Spectrum.}
 The $U(1)$ symmetry 
 \br
 A^{\pr} = e^{ia\b_0^2} A, \quad B^{\pr} = e^{-ia\b_0^2} B, \quad 
 c_k^{\pr} = e^{{-ia\b_0^2 }\o {n}}c_k
 \nonu
 \er
 of ${\cal L}_v$ (\ref{1.3}) gives rise to the following ``electric'' conserved
 current 
 \be
  J^{\mu ,vec}_{el } = i\( {{A \pa ^{\mu}B - B \pa ^{\mu }A}\o {1-AB}} + 
 \pa^{\mu } ln (c_1 c_2 \cdots c_{n-1} )\)
 \label{5.16}
 \ee
 
 As in the axial model ($J^{\mu ,ax}_{el} = 2\b_0 \eps^{\mu \nu } \pa_{\nu }R$) it can
 be realized in terms on the nonlocal field $u$ (or $\tilde u$) from eqns.
 (\ref{vecu}) and (\ref{5.11}), i.e.
  $ J^{\mu ,vec}_{el} = -i\eps^{\mu \nu } \pa_{\nu }ln (\tilde u^2)$.  The
  corresponding {\it electric charge } 
  \be
   Q_{el}^{vec} = \int_{-\infty}^{\infty} J^{0,vec}_{el}dx = 
  -i\int_{-\infty}^{\infty}dx \pa_x ln (\tilde u^2)
  \label{5.17}
  \ee
  {\it vanishes} for the 1-soliton solutions (\ref{5.14}), i.e. $ Q_{el}^{vec}
  =0$.  Observe that according to the axial-vector change of variables (\ref{2.23}) we have
  \br
  {\tilde u}^2 = {{\psi }\o {\chi }}. \quad \quad \tilde u = 
  {{u (c_1 c_2 \cdots c_{n-1})^{1\o 2}}\o {{(AB-1)}^{1\o 2}}}, \quad 
  \tilde u = e^{i \b_0 \theta }
  \nonu 
  \er
  and therefore the {\it electric current } $ J^{\mu ,vec}_{el} $ of the
  vector gauged model  is T-dual (see ref. \cite{ggsz1}) of the {\it
  topological current } $J^{\mu ,ax}_{\theta} = 
  -{i\o {2\b_0^2}}\eps^{\mu \nu }\pa_{\nu }ln({{\psi }\o {\chi }}) = {1\o
  {2\b_0^2}} J^{\mu ,vec}_{el}$ of the axial
  model (\ref{1.1}) (remember  that $Q_{\theta}^{ax} =0$ for the 1-soliton solutions
  of the axial model (\ref{4.6}) and (\ref{4.10})).  Next step is to recognize
  that the {\it b.c. } (i.e. $\a $) dependence of the vector model {\it soliton
  spectrum }
($  M^{vec},  E^{vec},  Q_{el}^{vec},  Q_{\tilde \theta }^{vec}, Q_{mag}, s^{vec} $) comes now from the 
topological
current( for $cos \a > 0 $)
\be
   J^{\mu ,vec}_{\tilde \theta} = -i\eps^{\mu \nu }\pa_{\nu } ln
  {A^2}, \quad ,\quad
   Q_{\tilde \theta}^{vec} =-i\int_{-\infty}^{\infty} dx \pa_x
  ln A^2 = {4n}  ( \a - {{\pi }\o {2}} Sign (\a )) + 4\pi  j_{\varphi}
  \label{5.18}
  \ee
  It turns out to be T-dual \cite{ggsz1} to the axial model {\it electric current}  
 $J^{\mu ,ax}_{el} = 2\b_0\eps^{\mu \nu } \pa_{\nu }R$.  It reflects the fact
 that the vector model fields $A$ and $B$ have nontrivial asymptotics at $x
 \rightarrow \pm \infty$, 
 \be
 A(+\infty ) = e^{in\a } = {1 \o {B(+\infty )}}, \quad 
 A(-\infty ) = e^{-in( \a - \pi Sign (\a ))  } = {1 \o {B(-\infty )}}
 \label{5.19}
 \ee
 The origin  of the symmetries  generated by  the current $ J^{\mu ,vec}_{\tilde \theta}$ is similar to
 one of the axial model (\ref{2.11}),
 \br
 A^{\pr }= e^{i\pi s_1} A, \quad B^{\pr }= e^{i\pi s_2} B,
 \quad c_k^{\pr }= e^{-i\pi {{s_1 }\o n}} c_k 
 \nonu 
 \er
 $s_1 + s_2 =2K, \;\;s_1 - s_2 =2L$.  One might find  contradictory  the
 continuous $\a$ dependence of the topological  charge $ Q_{\tilde \theta}^{vec}$
  (\ref{5.18}).
  In fact $\a$ should be {\it quantized}, i.e. $\a - {{\pi}\o 2}Sign (\a) = 
  {{\b_0^2}\o {4n}}(j_{el} - {{4\pi}\o {\b_0^2}}j_{\varphi})$,
   as we shall see in Sect. 7.
   
 The remaining (true) topological charges $ \tilde Q_k$ (i.e. the vector
 analogs of $Q_k^{top}, J^{\mu}_k = {{2n}\o  {\b_0 }}\eps^{\mu \nu} \pa_{\nu} \varphi_k $
 ),  can not be defined as $Q_k = \int dx \pa_x ln(c_k)$
 since the asymptotics of $c_k$  include certain $\a$ dependence, as one can
 see from eqn. (\ref{5.14}),
 \be
 c_k (+\infty) = e^{-i\a}, \quad c_k (-\infty) = e^{i( \a - \pi Sign (\a )) } 
 \label{5.20}
 \ee
 Taking into account the axial-vector transformation (\ref{2.23}) and the
 vector model 1-soliton  asymptotics (\ref{5.19}) and (\ref{5.20}) it is easy
 to check that 
 \br
 {{i} \o { \b_0 }}ln (c_k c_{k+1} \cdots c_{n-1} A^{{n-k}\o {n}} ) |_{\pm
 \infty} = {{2\pi }\o { \b_0 }}{k \o n} \tilde N_{\pm}, \quad k=1,2, \cdots,
 n-1
 \nonu 
 \er
 and $\tilde N_{\pm}$ are arbitrary integers.  Hence we can take
 \be
 \tilde Q_k^{top} = {{2n}\o { \b_0 }}\int_{-\infty}^{+\infty} \pa_{x} 
 ln (c_k c_{k+1} \cdots c_{n-1} A^{{n-k}\o {n}} )= {{4k \pi }\o {\b_0^2 }} 
 j_{\varphi } = Q_k^{top}
 \label{5.21}
 \ee
 It is important to note that although the zeros of the vector model potential
 (i.e. its vacua) are again labeled by two integers ($\tilde L, \tilde N $)
 the vector 1-solitons (\ref{5.14}) interpolating between two vacua with both  
 $\tilde L$ and $\tilde N$ {\it different from zero }, 
 $j_{ \tilde \theta}= \tilde L_+ - \tilde L_- \neq 0 $ (see (\ref{qeteta})) , $ j_{\varphi }=\tilde N_+ - 
 \tilde N_- \neq 0 $  
 contrary to the axial model charged solitons (\ref{4.6}) and 
 (\ref{4.10}) that have $Q_{\theta}^{ax} \sim L_+ - L_- = 0$, i.e. relates the
 trivial vacua $(0,0)$ to $(0, N)$ (for $\nu =0$, i.e. without  topological $\theta$-term).

 The derivation of the 1-solitons energy $ E^{vec}$ and momentum $ P^{vec}$ is similar to the
axial case presented in Sect. 3.  The main tools are again the soliton conservation laws
(\ref{5.10}) and (\ref{5.11}) that allow to show that $T_{01}^{vec}$ and $T_{00}^{vec}$ are
indeed total derivatives.  Thus, we find that the $ E^{vec}$ and $ P^{vec}$ are certain functions
of the asymptotics of fields $A,B ,c_k$  (\ref{5.19}) similar to the axial $E$ and $P$
(\ref{3.19}) and (\ref{3.21}) replacing $Q_{el}^{ ax}$ by its dual $ Q_{\tilde \theta}^{vec}$,
\be
 M^{vec} = {{4n m}\o { \b_0^2 }} |\sin {{1 }\o {4n}}( Q_{\tilde \theta}^{vec} - 
 \b_0^2   Q_{mag})| = M^{ax}
\label{5.22}
\ee

In order to complete the analogy with the axial model soliton spectrum (including $\nu$ and $ \tilde
\g$)  we have to add  a purely topological CPT-breaking 
 term to the CPT-invariant vector
Lagrangean (\ref{1.3}) ${\cal L}_{impr} = {\cal L}_{vec} + \delta {\cal L}_{top}^{v}$, 
\be
\delta {\cal L}_{top}^{v} = \sum_{k=1}^{n-1} {{\g_k}\o {\b^2}}
 \eps^{\rho \nu } \pa_{\rho} ln (c_k \cdots
c_{n-1}) \pa_{\nu} ln ({A\o B})
\label{5.23}
\ee

Although  the equations of motion remains unchanged  $\delta {\cal L}_{top}^{v}$ 
contributes to
$ J_{el}^{\mu, vec }$ (i.e. to $J^{\mu, ax}_{\theta}$) as it was explained in Sect.2.5 :
\br
 Q_{el,impr}^{vec} &=&  \b_0 \int _{-\infty}^{\infty} \pa_{x}\theta_{impr} dx = 
 Q_{\theta}^{ax} + 4\pi \tilde \g j_{\varphi}= Q_{\theta, impr}^{ax} \nonu \\
 Q_{\tilde \theta, impr}^{vec} &=& \b_0 \int _{-\infty}^{\infty}
 \pa_{x} \tilde \theta_{impr} dx = 
 Q_{el}^{ax} - {{\nu \b_0^2}\o {2\pi}} j_{\varphi} 
 \label{5.24}
\er
Taking into account that $ {{Q_{\tilde \theta , impr}^{vec}}\o {\b_0^2}}$ is a topological charge, i.e.
\be
{{2\pi}\o {\b_0^2}}  Q_{\tilde \theta , impr}^{vec} = 2\pi j_{\tilde \theta},
\quad  j_{\tilde \theta}
= 0, \pm 1, \pm 2, \cdots 
\label{qeteta}
\ee
 we find 
\br
 Q_{\tilde \theta}^{vec} = \b_0^2 (j_{\tilde \theta} + {{\nu}\o {2\pi}}j_{\varphi} ) = 
Q_{el}^{ax}(j_{el} \rightarrow j_{\tilde \theta})
\label{5.26}
\er
The complete discussion of the T-{\it duality relations} between the solitons and breathers
spectrum of the axial and vector models requires further investigation.

\sect{  Soliton Spin}

{\bf 6.1 Weak coupling spectrum.}
 The fields (and particles) in 2-d relativistic theories belong to certain 
 representations of the 2-d Poincare group, ${\cal P}_2 = O(1,1)
 \otimes T_2 $.  As one can see from its algebra:
 \br
 [M_{01}, P^{\pm} ] = \pm P^{\pm}, \quad \quad [P^{+}, P^{-}] =0
 \nonu 
 \er
 the Lorentz boosts $M_{01}$ and the mass operator  $2P^+P^-$ provide a 
 set of mutually commuting operators.  Their eigenvalues ($s,
 M^2$) are used to characterize the ${\cal P}_2$ representations.  For example, 
 the field $\phi_{(s)} (z, \bar z)$ of Lorentz spin $s$ tranforms under $O(1,1)$ 
 Lorentz rotations : $z^{\pr} = e^{\d} z, 
\bar z^{\pr} = e^{-\d} \bar z$ ($\d $ real ) as
\br
\phi^{\pr}_{(s)} (z, \bar z) = e^{\d M_{01}} \phi e^{-\d M_{01}}=
e^{s \d}\phi_{(s)} (z^{\pr}, \bar z^{\pr}) 
\nonu
\er
or infinitesimaly,
\br [M_{01}, \phi_{(s)} (z, \bar z) ] = 
\(z \pa - \bar z \bar \pa + s\) \phi_{(s)}(z, \bar z)
\label{a1}
\er
In ${\cal P}_2$ invariant theories $M_{01}$ appears as $O(1,1)$ Noether charge in 
the well known form $iM_{01} = \int_{-\infty}^{\infty}
xT_{00} dx$.  Taking the explicit form of $T_{00}$ (say, (\ref{3.15})) in terms of 
fields and their momenta and using  the  canonical commutation relations it is 
straightfoward to derive the {\it canonical spins} of all fields:
\br
s_{\varphi_{l}} = 0, \quad s_{\psi} = - s_{\chi} = {{(n-1)}\o 2}
\label{a2}
\er
Similarly we find for their $U(1)$ charges 
\br
Q_{el}^{\varphi_l} = 0, \quad Q_{el}^{\psi} =-Q_{el}^{\chi} = \b_0^2
\label{a3}
\er
It is instructive to write eqns. (\ref{a2}), (\ref{a3}) in a compact form, 
relating spins with electric charges:
\br
s = {{(n-1)}\o {2 \b_0^2}}Q_{el}
\label{a4}
\er
We complete the discussion of the spectrum of ``weak coupling'' 
($\b_0^2 \rightarrow 0$) fields (particles) by noting that the ``bare'' masses 
of the ``weak'' fields of the $A_n^{(1)}$ model (\ref{1.1}) are given by 
\br
m_{\psi} = m_{\chi} = m, \quad m_{\varphi_l} = 2m \sin ({{\pi l}\o {n}})
\er

{\bf 6.2 Strong coupling particles.}
The main characteristic of the ``weak'' coupling particles is that 
they do not carry topological charges, $Q_{\theta}=0, \;\;Q_{l}^{top}=0$, 
as one can easily check by substituting the weak coupling  ``free'' 
solutions into (\ref{jota}) and 
(\ref{carga}).  It reflects the fact that all ``weak'' fields have vanishing 
assymptotics at $x \rightarrow \pm \infty$.  The analysis of the classical
 vacua structure (i.e. nontrivial constant solutions) of the
axial $A_n^{(1)}$ model presented in Sect. 2.3 allows to list all the 
addmissible boundary conditions 
\br
\varphi_{l}^{(N)}(\pm \infty) = {{2\pi }\o {\b_0}} {{l N_{\pm}}\o {n}},
 \quad r(\pm \infty ) =0, \quad \theta^{(L)}(\pm \infty ) = {{\pi
L_{\pm}}\o {2 \b_0}}
\label{a6}
\er
The solutions of eqns. (\ref{2.1}) with nonvanishing asymptotics 
(\ref{a6}) such that $j_{\varphi} = N_{+} - N_{-}\neq 0$ and/or 
$j_{\theta}= L_+ - L_-\neq 0$ are by construction  topologically charged. 
The simplest example $j_{\varphi} =1, j_{\theta}=0$ of finite energy 
electrically charged topological soliton is given by eqns.
(\ref{4.6}) and (\ref{4.10}).  As it  is expected its semiclassical 
spectrum (\ref{1.5}) shows the characteristic ``strong coupling'' 
dependence on the coupling constant $\b_0^2$.  They 
define a ``strong coupling'' set of particles labled by 
$(M, s; Q_{el}, Q_{mag}, Q_{\theta})$.  The knowledge of the explicit values of
these conserved charges (classical, semiclassical and quantum ) 
is crucial in answering the question about ``strong coupling''
symmetry group of the model as well as  in the construction of the 
dual model in terms of the ``strong fields'' (carrying the strong
coupling particles quantum numbers ).  It is clear that the spin of 
the strong fields is the main ingredient 
in writing the kinetic part of the S-dual Lagrangean.

{\bf 6.3 Lorentz spin of electrically charged topological solitons.}
As we have shown in Sect. 3.2, the fact that 1-soliton stress-tensor is a total derivative
\br
T_{00} = \pa_x \( F^{-} - F^{+} \), \quad T_{01} = \pa_x \( F^{-} + F^{+} \) \nonu
\er
is crucial in the demonstration that solitons energy (and mass) get contributions 
from the boundary terms only.  One expects that similar
arguments take place in the calculation of the soliton spin.  What we need in this case is to 
represent $T_{00}$ (and $T_{01}$) as certain second derivatives
\br
T_{00} = -{{2}\o {\b}} \pa_x^2  \( \sum_{k=1}^{n-1}\varphi_k + {{n-1}\o {2}}R + n \zeta \), 
\label{a7}
\er
An auxiliary field $\zeta$ we  have introduced in (\ref{a7}) is defined as 
 solution of the following first order equations:
\br
\pa \zeta &=& {m \o \b }\g e^{\b (\varphi_{n-1} +{R\o n})}, \nonu \\
\bar \pa \zeta &=&-{m \o {\b \g }}\( e^{\b (\varphi_{1} -{R\o n})} + \b^2 \psi \chi e^{-\b {{R}\o {n}}}\)
\label{a8}
\er
As a consequence of (\ref{a8}) and eqns. (\ref{3.61a})-(\ref{3.8})  we realize that
 $\pa \bar \pa \zeta $ is proportional to the trace $T_{\mu}^{\mu}$
 of the stress-tensor $T_{\mu \nu}$ \footnote{the field $\tilde \zeta$ with the above properties 
 together with the free field $\eta$ are
 familiar from the conformal affine extension \cite{luiz} of the affine NA-Toda models } :
 \br
 \pa \bar \pa \tilde \zeta = T_{\mu}^{\mu} = {m^2\o \b } \( e^{\b (\varphi_1 + \varphi_{n-1})} 
 + \b^2 \psi \chi e^{-\b \varphi_{n-1}}\), \quad 
 \tilde \zeta =  \zeta + {m^2 \o \b} z \bar z
 \label{a9}
 \er
 The proof of eqn. (\ref{a7}) is based on the following identity
 \br
 \pa_x \(\sum_{k=1}^{n-1} \varphi_k + {{n-1}\o 2} R + n \zeta \) = {\b \o 2} \( F^+ - F^- \)
 \label{a10}
 \er
 which can be easely checked taking into account eqns. (\ref{3.61a})-(\ref{3.8}) and (\ref{a8}).  With the explicit 
 1-soliton solutions (\ref{4.5})-(\ref{4.7}) at
 hand we find that $\zeta$ satisfying eqns. (\ref{a8}) has the form
 \br
 \zeta = {1\o \b} ln \( 1+ e^{2f}\) - 2{m\o \b} (\cos \a ) \rho_+ + 2i {m\o \b}  (\sin \a )\rho_-
 \label{a11}
 \er
Note that for imaginary coupling $\b = i \b_0$ the stress tensor is complex, say for $T_{00}$ we have
\br
T_{00} = {2\o \b_o^2} \pa_x^2 \tau, \quad \tau = i \b_0 \( \sum_{i=1}^{n-1}  \varphi_k + {{n-1}\o 2}R  + n  \zeta \)
\label{a12}
\er
Although the energy (\ref{3.19}) of the 1-solitons is real and finite, their spin $M_{01}$ is in general complex and also contains an
infinite part
\br
M_{01}= -i \int_{-\infty}^{\infty} x T_{00} dx = -{{2i}\o {\b_0^2}} \(x \pa_x \tau - \tau \) |_{-\infty}^{\infty} \nonu
\er
Its real part, however is finite
\br
s = Re \( M_{01} \)  = {{(n-1)}\o {2 \b_0^2}}\( Q_{el}^{ax} + \b_0^2 Q_{mag} \)
\label{a13}
\er
and we take it as a definition of the spins of 1-solitons in consideration.  It is worthwhile
 to mention that eqns. (\ref{a7})-(\ref{a13}) with $R=0$, ($Q_{el} =0$) remain valid for the solutions of the $A_{n-1}^{(1)}$-abelian
 affine Toda, i.e. the spins of the 1-solitons of this model turns out to be
 \br
 s_{Ab. Toda }= {{n-1}\o  2}Q_{mag}
 \label{a14}
 \er

\sect{Towards Exact Quantization}
{\bf 7.1 Semiclassical Quantization.}
Classical periodic motions are known to correspond to quantum bound states (discrete energy
spectrum $E_j$) such that the frequency of the emition $E_j \rightarrow E_{j+1}$ coincides
with the frequency of the classical motion.  The energy eigenvalues $E_j$ can be found from the
Bohr-Sommerfeld quantization rule $\int_{0}^{\tau } dt p \dot {q} = 2\pi j$ ($j$ an integer ),
i.e.
\be
E_j \tau + S = 2 \pi j , \quad \quad S = \int _0^{\tau } dt {\cal L}
\label{6.1}
\ee
As we have shown in Sect. 4 (see eqn. (\ref{4.13})) our charged 1-solitons (\ref{4.6}),
(\ref{4.10}) at the rest frame $v= \tanh (b)=0$ represent periodic particle-like
 motion $\tau =
{{2 \pi }\o {m\sin\a}}$ similar to the SG breather.  Hence, the {\it semiclassical }
 soliton
spectrum can be derived from the field theory analog    of eqn. (\ref{6.1})
\br
&&{\cal H} = \Pi_{\rho_i} \dot \rho_i  - {\cal L}, \quad \Pi_{\rho_i} = {{\d {\cal L}}\o {\d
\dot \rho_i}}, \quad \rho_i = (\psi , \chi, \varphi_k)
 \nonu \\
&&S + E(v=0) \tau = \int_0^{\tau } \int_{-\infty}^{\infty} \Pi_{\rho_i} \dot \rho_i 
= 2\pi j_{el}
\label{6.2}
\er
Taking into account the relations between $\psi \Pi_{\psi} - \chi \Pi_{\chi}$ and $\pa_xR$
we realize that 
\br
\Pi_{\psi} \dot \psi + \Pi_{\chi} \dot \chi = -{{i}\o {\b_0}}\pa_x R \pa_t ln({{\chi}\o
{\psi}}) + {{i}\o {\b_0}}\pa_t R \pa_t ln(\chi \psi)
\nonu
\er
At the rest frame ($\cosh b = 1,\;\; \sinh b =0$) our 1-soliton solutions 
(\ref{4.6}), (\ref{4.7}) and (\ref{4.9}) have a simple t-dependence:
\br
\pa_t ln ({{\chi}\o {\psi}}) = 2im \sin \a, \quad \pa_t ln \psi \chi = 0, \quad \pa_t
\varphi_i = \pa_t \Pi_{\varphi_i} =0 
\nonu 
\er
Therefore we find (for $\nu =0$) 
\br
\int_{0}^{\tau} \int_{-\infty}^{\infty} \Pi_{\rho_i}\dot \rho_i = {{2\pi }\o {\b_0^2}}Q_{el}^{ax}
\nonu
\er
In the case $\nu \neq 0$ the corresponding  momenta of the fields $\Pi_{\rho_i}$ acquire
improvements resulting in
\br
\int_{0}^{\tau} \int_{-\infty}^{\infty} \Pi_{\rho_i}\dot \rho_i = {{2\pi }\o {\b_0^2}} \int
_{-\infty}^{\infty} J^{0, ax}_{el, impr} dx
\nonu 
\er
This leads to the following semiclassical quantization of $Q_{el}$ (see eqn. (\ref{jelimpr})
for the definition of  $J^{0, ax}_{el, impr}$):
\br
Q_{el}^{ax} = \b_0^2 ( {{\nu }\o {2\pi}}j_{\varphi} + j_{el})
\label{quant}
\er

We find instructive to present an alternative derivation of 
the quantization rule (\ref{6.2}), (\ref{quant}) by
explicit calculation of 
the action $S(\tau ) = \int_0^{\tau } \int_{-\infty}^{\infty} dx {\cal
L}_{a} $ on the 1-soliton solutions.  The key point is to demonstrate that ${\cal L}_a$ is 
 a
total derivative due to the first order equations (\ref{3.61a}) and (\ref{3.11}) and the soliton
conservation laws (\ref{3.12}), (\ref{3.13})  and (\ref{3.14}).  The  calculation is a bit more
involved than the similar one in (\ref{3.15}) concerning $T_{00}$ and $T_{01}$.   That is why we
consider  the simplest case of $A_2^{(1)}$ only ($\g =1, \cosh (b) =1, v=0, \nu =0$)
\br
{\cal L}_a = \pa \varphi \bar \pa \varphi + {{\bar \pa \psi \pa \chi }\o {\Delta }}e^{-\b \varphi }
-\b^2V_a = -T_{00} + {2m \o \b} \sin \a \pa _{x} R
\nonu
\er
where we have used 
\br
\pa R = {2\o {y\b }}(e^{-\b R} + e^{\b R} -2 - y^2), \quad \quad 
\bar \pa R = -{2\o {y\b }}(e^{-\b R} + e^{\b R} -2 - y^2)
\nonu
\er
Therefore we have,
\be
S_{M}(\tau ) = - \tau E(v=0) + {{2\pi }\o {\b_0^2}} Q_{el}^{ax},
\label{6.7}
\ee
that together with (\ref{6.2}) leads again to (\ref{quant}).  It is important to note that
repeating this calculation for the Euclidean case $t \rightarrow it$ we find 
\be
S_E (\tau ) = {{2\pi }\o {\b_0^2}} Q_{el}^{ax}
\label{6.8}
\ee
i.e. the Euclidean charged 1-solitons (they are not static ) have properties similar to the
instantons, i.e. the Euclidean action is bounded below by the electric charge.  In the
 case of
the  vector gauged model (\ref{1.3}) all the results have the same form as above with 
$Q_{el}^{ax} \rightarrow Q_{\tilde \theta}^{vec}$.

{\bf 7.2 Exact Counterterms and $\b$ renormalization.}
The quantum properties of the $A_n^{(1)}$ dyonic integrable modes (\ref{1.1}) and (\ref{1.3})
 turns out to be quite similar to the CSG model \cite{devega}.  While the {\it main effect }  of
 the soliton quantization of the SG and  $A_n^{(1)}$ abelian affine Toda theories is the
 coupling constant renormalization \cite{holl},\cite{dash}  in the CSG, as well as in the models in
 consideration the  1-loop calculation shows that new counterterms are needed in order to
 define a consistent quantum theory.  
 
 The same phenomenon takes place in the ``free limit'' $V=0$ ( i.e. $m=0$) of 
 (\ref{1.1}) and (\ref{1.3})  known to describe strings in 
 curved background: 2-D blackhole
 ($n=1$), 3-D black string ($n=2$), etc \cite{horne}. 
 Since they can be realized as gauged
 ${{SL(2) \otimes U(1)^{n-1}} \o {{U(1)}}}$ WZW models, 
  the best way \footnote{ an
 alternative way is to use the representation theory of the parafermionic 
 extensions of the
 Virasoro algebra.  The quantum theory of the conformal `` non free'' 
 $A_2^{(1)}$ NA-Toda
 model is shown to be equivalent to the representation theory of the $V^{(1,1)}_{3}$-algebra
 \cite{annals}}  of deriving  the corresponding quantum
  theories (i.e.
 renormalization, counterterms, etc ) is by applying the functional integral 
 methods \cite{tsey}.
  The path integral formalism for the conformal limit of our dyonic models 
  (i.e. $V_{def} =0$ 
  in
  (\ref{1.2})) \cite{annals} was extended in our recent paper \cite{ggsz2} to the case of a 
  large family
  of integrable models, that can be represented as two loop gauged WZW model $\hat H_{-}
  \backslash G_n^{(1)} /\hat H_+ = G_0 /U(1) , \;\; G_0 = SL(2) \otimes U(1)^{n-1}$. 
  Integrating over the infinite $\hat H_{\pm}$ but keeping $U(1)$ (spanned by $\l_1 \cdot
  H^{(0)}$)
  ungauged, we find the following effective action (of the intermediate model 
  (\ref{1.4}) for
  $A_0 = \bar A_0=0)$
  \br
  S_{eff}^{G_0} &=& S_{WZW}^{G_0} (D_v) - {{1}\o {\b^2_{0,ren}}}\int dx^2 Tr \(\eps_+ D_v
  \eps_- D^{-1}_v \) \nonu \\
  &+ &
  {1 \o {\b^2_{0,ren}}} \int dx^2 Tr \( A_0 \bar \pa D_v D_v^{-1} - \bar A_0 D_v^{-1} \pa D_v -
  D_v^{-1} A_0 D_v \bar A_0 + (1 + \l_0) A_0 \bar A_0 \) \nonu \\
  \label{6.9}
  \er
  Extending Tseytlin's arguments (see Sect. 3 and 4 of the first of refs 
  (\cite{tsey})) to the
  $S^{G}_{eff}$, given by (\ref{6.9}), we realize that 
  \br
  \b^2_{0,ren} &=& {{2\pi }\o {k-n-1}}, \quad \quad \b^2_0 ={{2\pi }\o {k }} \nonu \\
  \l_0 &=& 2{{n+1} \o {k-n-1}} = {{(n+1)}\o { \pi}}\b^2_{0,ren}
  \label{6.10}
  \er
   ( and $1 + \l_0 = 1, i.e. \l_0 = 0$ is the unrenormalized `` classical'' coefficient
   of $A_0 \bar A_0$ term responsible for the  cancellation  of the $U(1)$ gauge  anomaly). 
   Integrating out $A_0, \bar A_0$ in the partition function 
   \br
   Z^{v}_{A_n^{(1)}} = \int {\cal D}A_0 {\cal D} \bar A_0 {\cal D} D_v
   e^{-S_{eff}^{G_0}} = N_{\infty} \int {\cal D} D_v e^{-S_{eff }^{G_o /U(1)}}
   \nonu
   \er
   we find the renormalized quantum action for the dyonic model (\ref{1.3})
   \br
   S_{eff }^{G_0 /U(1)} = \int {\cal L}_v d^2x + {{n+1}\o {8\pi}} \int {{(A\pa_{\mu }B
   - B \pa _{\mu }A )^2}\o {(1- AB) (1 + \l_0 - AB )}} + i\int {{R^{(2)}}\o {\b_0
   }}\sum_{k=1}^{n} \b_{k} ln (c_k )d^2x
   \label{6.12})
   \er
   It differs from the classical action (\ref{1.3}) by 
   
   {\bf a)} the counterterm 
   \be
   {\cal L}_{c.t. } ={{n+1}\o {8\pi }}  {{(A\pa_{\mu }B
   - B \pa _{\mu }A )^2}\o {(1- AB) (1 + \l_0 - AB )}},
   \label{6.13}
   \ee
   
   {\bf b)} the  dilaton contribution 
    \be
    {\cal L}_{dil} = {{i}\o {\b_0}} R^{(2)} \sum_{k=1}^{n} \b_k ln c_k, \quad ln c_n =
    {{i\b_o}\o 2}ln (1-AB), \quad \b_1 = -3, \b_k = -1, k=2, \cdots ,n
    \label{6.13a}
    \ee
    ( $R^{(2)}$ is the worldsheet curvature).
    
    {\bf c)} the $\b $ renormalization
    (\ref{6.10}).
    
  Expanding denominators in (\ref{6.13}) we find it consistent
    with the 1-loop (in $\b^2_{0,ren}$) result (see ref.  \cite{devega} for $n=1$
    case and \cite{fat} for the $B_n^{(1)}$ T-self dual Fateev's models ).
    
    {\bf 7.3 $U_q(\hat {SL}(n+1))$ Symmetries of the Soliton Spectrum.}
    The most important question concerning the quantum solitons and breathers
    spectrum is about the solitons {\it symmetry group}.  Following the
    paralel with the $A_n^{(1)}$ abelian affine Toda models where the quantum 
    solitons are related to the representations with $q^{k+n+1}=1$ of the
    affine quantum group $U_q(\hat {SL}(n+1))$ \cite{bernard} one expects that neutral
    and charged 1-solitons of our dyonic models to have similar properties whether 
    it is true   is an open
    question, although there exist few hints that this is indeed the case:
    \begin{itemize}
    \item Neutral  1-solitons and their conservation laws coincides with the 
    $A_{n-1}^{(1)}$ abelian affine Toda solitons, that is why neutral  sector
    should manifest $U_q(\hat {SL}(n))$ symmetry  (but not $\hat {SL}(n+1, q)$).
    
    \item The form of the classical conservation laws (\ref{3.12}) for the
    charged solitons is similar to the abelian case and 
    the  classical "braiding relations" of the currents components encoded in
    the classical $r$-matrix  \cite{dubna} are supporting the
    conjecture the $U_q(\hat {SL}(n+1))$ is the charged solitons group of
    symmetries.
    
    \item  The $n=1$ case (i.e. the quantun CSG model ) corresponds to certain
    thermal perturbations  of the parafermionic conformal models \cite{fatdevega}.  Its
    quantum S-matrix  confirms the
     $U_q(\hat {SL}(2))$  symmetry in this case.
     
     \item The $n=2$ (i.e. $A_2^{(1)}$) dyonic model can be realized as an
     appropriate perturbation of the $V_3^{(1,1)}$-algebra conformal minimal
     models constructed in \cite{annals} (see Sect. 9 of \cite{annals}).  Our preliminary
     calculation based on  bosonization of the vertices representing
     quantum solitons shows that the $A_2^{(1)}$ dyonic models possess a 
     nontrivial set of quantum conserved currents, whose  charges span the
     $SL(3)_q$ quantum affine algebra.
  \end{itemize} 
\sect{Concluding Remarks}  
{\bf 8.1 S-duality.}  
The particle-like {\it nonperturbative} solutions of $SU(n+1)$ YMH model-
monopoles and dyons are believed to provide the fields degrees of freedom
relevant for the description of the {\it strong coupling phase } of YMH (and of
QCD in general ).  Similar phenomena is known to take place in 2-D IM:  Topological
solitons serve as {\it strong coupling}  variables {\it dual } to the fundamental ({\it weak
coupling }) particles (fields), presented in the IM's  `` weak coupling '' actions (say,
(\ref{1.1})) \cite{col},\cite{fat}.  The simplest and best understood example is the strong-weak 
 coupling  (S-) duality   between the massive Thirring and the sine-Gordon models \cite{col},\cite{man}. 
 
 The knowledge of the exact {\it quantum dyons } spectrum is crucial in 
 the derivation of
 the YMH model {\it strong coupling  symmetry group} and its representations to be used in the
 construction of the strong coupling YMH effective action.  The Montonen-Olive {\it
 duality conjecture} \cite{mont} states that the monopoles of the $G_n$-YMH model belongs to
 fundamental representations of the dual group $G_n^v$ (say, $G_n= SU(n+1)$, and
 $G_n^v=SU(n+1)/Z_{n+1}  $).  It has been proved for $N=4$ SUSY $G_n$-YM theories 
  and certain $N=2$ SUSY YM-matter models \cite{vafa}.
 
  For $N=1$ and
 for the {\it non-SUSY} YMH models, the exact dyonic spectrum is {\it unknown}.  As it was
 argued in the introduction, the study of the $U(1)$ {\it charged topological 
  solitons} of appropriate 2-D IM of dyonic type and their exact quantization  may
  contribute  to
  the understanding  of the nature of strong coupling symmetry groups of non SUSY YMH
  theory.  The conjectured form of the exact quantum 2-D dyons spectrum and the arguments presented in sect.7.3
  makes feasible the role of the centerless
  quantum affine group $A_n^{(1)}(q)$, ($q = e^{i \hbar \b_{0,ren}^2}, \;\;
  \b^2_{0,ren} = {{2\pi }\o
  {K-n-1}})$ as strong coupling symmetry group of the $A_n^{(1)}$-dyonic 
  IM's (\ref{1.1}) and
(\ref{1.3}).  The question  of whether certain affine  {\it quantum group} appear in the
description of the 4-D YMH quantum dyons and domain walls is far from being answered.  The observation
concerning the affine $\hat {SU}(n+1) $ symmetries of the classical solutions of the 
$ {SU}(n+1) $ self dual YM (and the YMH-Bogomolny) equations \cite{dolan2} is an indication that their
finite energy quantum solutions  could have    the
 $ U_q(\hat{SU}(n+1)) =U_q(A_n^{(1)}) $  as an algebra of symmetries.  The problem to be solved
 before any attempt for {\it affine quantum group improvement} of the original
  Montonen-Olive
 conjecture is about the {\it integrable models} S-dual to the $A_n^{(1)}$ dyonic models (\ref{1.1}) and 
 (\ref{1.3}).  The fundamental fields used in writing their Lagrangeans have to carry the
 quantum numbers ($Q_{el}, Q_m, M, s, \l_j, \cdots $) of charged and neutral solitons of
  the $A_n^{(1)}$ models.  Its soliton
 (i.e. strong coupling ) spectrum should have the quantum numbers of the fundamental
 fields of the dyonic models i.e.$\psi,\chi,\varphi_i$.  Similarly to the solitons of all
  known 2-D integrable models
 they must have certain affine quantum group structure behind. The most natural candidate is the
 algebra dual to  $A_n^{(1)}(q)$ with $\tilde q= e^{i\hbar \tilde \b^2}, \; \; 
 \tilde \b^2 = {{1\o { \b^2}}}$.  The  partition functions of such pair of S-dual IM's
 are expected to be related by the S-duality transformations 
 \br
 \tau \longrightarrow {{n_1 + m_1 \tau }\o {n_2 + m_2 \tau }}, \quad \tau = {{\nu }\o
 {2\pi }} + i {{2\pi }\o {\b^2_r}}
 \nonu 
 \er
 where $\twomat{n_1}{m_1}{n_2}{m_2} \in SL(2,Z)$.  We have {\it not a recipe} of how to construct the S-dual of a
 given IM satisfying all the requirements listed above.  An important experience in this
 direction are the families of S-dual pairs of IM introduced and studied by Fateev \cite{fat}
 \br
 B_{n+1}^{(1)} (\{ \chi , \bar \chi , \phi_i \}_{strong}, \b^2  ) & \rightarrow   &
 A_{2n+1}^{(2)} (\{ \psi , \bar \psi , \varphi_i\}_{weak}, {{4\pi} \o {\b^2}}  )\nonu \\
 A_{2n}^{(2)} (\{ \chi , \bar \chi , \phi_i \}_{strong}, \b^2  ) &\rightarrow &  
 A_{2n}^{(2)} (\{ \psi , \bar \psi , \varphi_i\}_{weak}, {{4\pi} \o {\b^2}}  )\nonu \\
 D_{n+1}^{(2)} (\{ \chi , \bar \chi , \phi_i \}_{strong}, \b^2  ) &\rightarrow   &
 C_{n}^{(1)} (\{ \psi , \bar \psi , \varphi_i\}_{weak}, {{4\pi} \o {\b^2}}  )
 \er
 The ``strong coupling models'' $G_n^{(a)}$ represent the complex SG (Lund Regge) model
 ($\chi, \bar \chi $)  interacting with $G_{n-1}^{(a)}$ abelian affine Toda theories
 ($\phi_i$).  Their actions are similar to (\ref{1.3}) and as it is shown in
 \cite{ggsz2} they can be derived following the general Hamiltonian reduction 
 procedure (or from gauged two loop WZW model) discussed in Sect. 2.  The explicit form of
 the $\lie_n^{(1)}$ valued flat connections depend on the specific choice of the grading
 operator $Q$, constant elements $\eps_{\pm}$ and in the way the chiral $U(1)$ symmetry
 $\lie_0^0$   is gauged fixed. 
The ``weak coupling models''  $\tilde G_n^{(a)}$ represent massive Thirring
 fermions  $\psi, \bar \psi $ interacting with the abelian affine Toda model
based on the dual algebra $\lie_{n-1}^{v}$ of $\lie_{n-1}^{(1)}$, i.e. 
 $ B_{n+1}^{(1)} $ is dual to $A_{2n-1}^{(2)}$, 
  $ D_{n+1}^{(2)} $ is dual to $C_{n}^{(1)}$, etc.   It is important to note
  that only the case of {\it real coupling} constant has been considered in ref.\cite{fat}. 
  Therefore, the corresponding potentials admit only trivial zero $\phi_i =
  0$ (but {\it not} $\phi_i \in {{2\pi }\o \b}\Lambda_{\omega}, \;\;
   \Lambda_{\omega}$ is the coroot lattice of $G_{n-1}^{(a)}$).  As a
   consequence, their finite energy nonperturbative solutions are the 
  $G_{n}^{(a)}$ analogs of the $U(1)$ charged {\it nontopological } solitons
  of the CSG model and their quantization is known to be related to
  $U_q(SL(2))$ ( and not to  $G_{n}^{(a)}(q)$).   This explains the pair of
  Thirring fermions representing these solitons  in the weak coupling
  model.  Whether the methods for quantization of these models, for
  construction of their exact S-matrix, etc.\cite{fat} can be applied to the imaginary
  coupling constant case is an {\it open question}.  Due to more complicated
  soliton spectrum of both models for $\b \rightarrow i\b_0 $ their S-duality
  for imaginary couplings has to be reconsidered.  Although the S-duality is
  a property of the quantum IM (i.e. counterterms, renormalization, etc.  are
  essential), an important {\it problem } to be addressed is whether exists
  an algebraic recipe ({\it the 2-d analog of Montonen-Olive conjecture}) that for
  any given IM $\{ G_n^{(1)}, Q , \eps_{\pm }, \lie_0^0 \}$ prescribes its
  S-dual IM (the classical limit only) 
  $\{ \tilde G_n^{(1)}, \tilde Q, \tilde {\eps}_{\pm }, \tilde {\lie}_0^0
  \}$.   The simplest problem to be solved is to find the {\it graded
  structure } behind the `` weak coupling '' Fateev's models  for real $\b $ and to
  construct their zero curvature representations.  The presence of the
  Thirring fermions interacting with the abelian Toda bosons is an indication
  that they are from the family of $G_{n}^{(1)\small{V}}$(the group dual of
  $G_{n}^{(1)})$ non abelian Toda-matter
  models of higher grade, say $\eps_{\pm } $ to be of grade $\pm 2$ (or some
  halfinteger grade ) and the physical fields to lie in the lowest grades
  $|s| <2$, as for  the IM's constructed in  ref. \cite{ferr}.
  
  {\bf 8.2 T-duality.}
   As we have mentioned in Sects. 1,2 and 5 the two $A_{n}^{(1)}$ dyonic
   IM's (\ref{1.1}) and (\ref{1.3}) studied in the present paper are
   T-dual by construction.  The abelian T-duality transformation
   (\ref{2.23}) and (\ref{func}) is specific for the models with target
   space metrics admitting one isometric coordinate $\theta = -{i\o {2\b_o}} ln
   ({{\tilde \chi }\o {\tilde \psi }})$ for (\ref{1.1}) and $\tilde
   \theta ={2\o{i \b_o}} ln (A)$ for (\ref{1.3}), (i.e. our models are invariant
   under global $U(1), \;\; \theta \rightarrow \theta + a,\;\;  
  \tilde \theta \rightarrow  \tilde \theta + \tilde a$).   The particular
  canonical transformation  $(\theta , \Pi_{\theta }) \rightarrow
  (\tilde \theta , \Pi_{\tilde \theta })$
  \be
 \Pi_{\theta } = -\pa_{x}  \tilde \theta, \quad \quad 
 \Pi_{\tilde \theta } = -\pa_{x}\theta
 \label{6.15}
 \ee
 which integrated form is eqn. (\ref{2.23}), are familiar from  the {\it
 string T-duality} \cite{giveon} relating certain {\it conformal } $\sigma $-
 {\it models } representing different curved string backgrounds of 3-D
 blackstring type \cite{horne}.  As it was pointed out in Sect. 2, the free
 limits $m=0, \; i.e. V=0$ of our dyonic models are nothing but the
 conformal $\s$-models considered in ref. \cite{horne} in a specific
 parametrization ( of Gauss type ) of the $A_n$ group elements and
 without the `` counterterm'' contributions (\ref{6.12}) and (\ref{6.13})
 essential for the exactness (in $\b^2$ ) of the`` target-space '' conformal metrics
 ($g_{ij}, b_{ij}, \phi_{dil}$).  The important difference is that in
 this $m=0$ case there exists more isometries $\varphi_i^{\pr} = \varphi_i
 + a_i $, together with $\theta^{\pr} = \theta + a_{\theta}$.  Adding
 the conformal ( and nonconformal ) vertices representing nontrivial
 (bounded  for $\b \rightarrow i \b_0$ ) potentials results in breaking
 $U(1)^{n} $ to $U(1)$, i.e. the ``free'' conformal T-duality group
 $O(n, n|Z)$ is broken to $O(1, 1|Z)$.  An important and {\it new}
 feature of the {\it non - conformal} T-duality specific for the dyonic IM's is
 the relation  between their soliton (and breathers ) solutions and the
 interchanges $Q_{el}^{ax} \rightarrow  Q_{\tilde \theta}^{vec}$ and 
  $Q_{\theta}^{ax} \rightarrow  Q_{el}^{vec}$ that are IM generalization of
  the well known maps between the momenta and winding numbers in the
  conformal (string ) case.  The complete discussion of the abelian
  T-duality for a large class of $U(1)$ symmetric axial and vector IM of
  dyonic type including the formal proofs of eqns. (\ref{2.23}) and 
(\ref{func})   are  presented in our work \cite{ggsz1}.

Since the T-dual IM's ( \ref{1.1}) and (\ref{1.3}) are expected to be
appropriate symmetry reductions of the 4-D $SU(n+1)$-YMH models  and
their (charged) topological solitons to be related to certain charged domain 
walls solutions,   the natural question  is about the 4-D
consequences of the T-duality observed in  2-D soliton spectra (see Sect.5).
It is clear that they have to represent specific residual (discrete) gauge
symmetries as in the string case \cite{giveon} reflecting the independence of the
strong coupling spectra of the manner of the local (unbroken) $U(1)$   gauge symmetry is
fixed.

The problem of the S and T-dualities of the soliton spectra of 2-D IM's of
dyonic type deserves special attention and more complete  investigation. 
  
{\bf Acknowledgments}  We are grateful to FAPESP,UNESP and CNPq for 
financial support.


\begin{thebibliography}{99}

\bibitem{ward} R.S. Ward, Phil. Trans. R. Soc. Lon. {\bf A315} (1985) 451; Lect. Notes Phys.
{\bf 280} (1987) 106; Lon. Math. Soc. Lec. Notes Ser. {\bf 156} (1990) 246

\bibitem{ablo}M.J. Ablowitz, S. Chakravarty and L. Takhtajan, PAM Report {\bf 108}(1991); {\bf
113}(1991); \CMP{158}{1993}{289};
 T.A. Ivanova and A.D. Popov, Theor and Math. Phys. {\bf 102}
(1995) 280;

\bibitem{leznov}  A. N. Leznov and M. V. Saveliev, Group Theoretical Methods for 
Integration of
Nonlinear Dynamical Systems, Progress in Physics, Vol. 15 (1992), Birkhauser
Verlag, Basel
\bibitem{ganoulis}N. Ganoulis, P. Goddard, D. Olive
 Nucl. Phys. {\bf B205},(1982), 601
 \bibitem{witt5}E. Witten, \PRL{38}{1977}{121}
\bibitem{shiff} A. Kovner, M. Shiffman and A. Smilga, Phys. Rev. {\bf D56}, (1997),7978-7989, also 
 G. Dvali and M. Shiffman, Phys. Lett. {\bf B396},(1997),  hep-th/9612128;
A. Hanany and K. Hori, \NPB{513}{1998}{119} hep-th/9706089; E. Witten, \NPB{507}{1997}{658}




\bibitem{gau}J. Gauntlett, D. Tong, P.  Townsend, 
 ``Supersymmetric intersecting domain walls in massive hyperkahler
 sigma models'', DAMTP-2000-69, hep-th/0007124;
 J. Gauntlett, R. Portugues, D. Tong, P.  Townsend, ``D-brane solitons 
 in supersymmetric models'', 
 DAMTP-2000-68, hep-th/0008221; 
 


\bibitem{bernard} D. Bernard and A. LeClair, \CMP{142}{1991}{99}


\bibitem{dolan2}L.Dolan,\PR{109}{1984}{1};
Ling-Lie Chau, Ge Mo-Lin and Wu Yong-Shi, \PRD{25}{1982}{1086}



\bibitem{lund1}F. Lund, \PRL{38}{1977}{1175};
B.S. Getmanov, JETP Lett. {\bf 25}(1977)119



\bibitem{dorey}N. Dorey and T.J. Hollowood, \NPB{440}{1995}{215}

\bibitem{devega} H.J. de Vega and J.M. Maillet, \PRD{28}{1983}{1441}

\bibitem{pousa}C.R. Fernandez-Pousa, M.V. Gallas, T.J. Hollowood and J.L. Miramontes,
\NPB{484}{1997}{609}; \NPB{499}{1997}{673};
C.R. Fernandez-Pousa and J.L. Miramontes,\NPB{518}{1998}{745}
\bibitem{annals}
J.F. Gomes,  G.M. Sotkov and A.H. Zimerman, 
  Ann. of Phys.{\bf 274}(1999)289, hepth/9803234; 
 Phys. Lett. {\bf 435B} (1998) 49, hepth/9803112,

\bibitem{ggsz1}J.F. Gomes, E.P. Gueuvoghlanian, G.M. Sotkov and
 A.H. Zimerman,``Torsionless T Selfdual Affine Non Abelian  Toda Models'',.
hep-th/0002173, To appear in the Proc. of the VI International 
Wigner Symposium, Istambul, Turkey, 2000; see also J.F. Gomes, E.P. Gueuvoghlanian, G.M. Sotkov and
 A.H. Zimerman, ``T-duality of axial and vector dyonic integrable models'',
 hep-th/0007116, to appear in Ann. Phys, (2001)
 
\bibitem{poly} A.M. Polyakov, Int. J. Mod. Phys. {\bf A5}(1990)833;
M. Bershadsky, \CMP{139}{1992}{71}

 
\bibitem{olive1} D. Olive, N. Turok and 
J. Underwood \NPB{409}{1993}{509}; \NPB{401}{1993}{663}

\bibitem{babelon}O. Babelon and D. Bernard, 
Int. J. Mod. Phys. {\bf A8}(1993)507; \CMP{149}{1992}{297}

\bibitem{luiz} L.A. Ferreira, J.L. Miramontes and J.S. Guillen, \NPB{449}{1995}{631}


\bibitem{zamora}L. Alvarez-Gaum\`e and F. Zamora, `` Duality in Quantum Field Theory (and String
Theory)'', hepth/9709180, Workshop on Fundamental particles and Interactions, Vanderbilt
University and CERN-La-Plata-Santiago-de-Compostela School of Physics, May 1997

 
\bibitem{gauntlett}J. Gauntlett, N. Kim, J. Park, P. Yi, \PRD{61}{2000}{125012}

\bibitem{ggsz2}
J.F. Gomes,  E.P. Gueuvoghlanian, G.M. Sotkov and A.H. Zimerman, 
``Dyonic Integrable Models'', hepth/0011187, to appear in Nucl. Phys. {\bf B}, (2001)
 


\bibitem{tsey}A. A. Tseytlin, \NPB{399}{1993}{601}; \NPB{411}{1994}{509}

\bibitem{holl}T.J. Hollowood, \NPB{384}{1992}{523}



\bibitem{liao}H.C. Liao, D. Olive and N. Turok, \PLB{298}{1993}{95}

\bibitem{aratyn} H. Aratyn, L.A. Ferreira, J.F. Gomes and A.H. Zimerman,
Phys. Lett {\bf B 254} (1991) 372; 
L.A. Ferreira, J.F. Gomes, A. Schwimmer  and A.H. Zimerman, \PLB{274}{1992}{65}

\bibitem{horne}J.H. Horn and G.T. Horowitz, \NPB{368}{1992}{444};
P. Ginsparg and F. Quevedo, \NPB{385}{1992}{527}

\bibitem{giveon}A. Giveon, M. Porrati and E. Rabinovici, \PR{244}{1994}{77};E. Alvarez, L. Alvarez-Gaum\`e and Y. Lozano, 
Nucl. Phys. Proc. Suppl. {\bf 41}(1995)1







\bibitem{witten1}E. Witten, \PLB{86}{1979}{283}
\bibitem{polch}J. Polchinski, String Theory, Vol. I,   Cambridge Univ. Press (1998)

\bibitem{dash}R.C. Dashen, B. Hasslacher and A. Neveu, \PRD{10}{1974}{4114};
 \PRD{10}{1974}{4130};  \PRD{10}{1974}{4138};  \PRD{11}{1975}{3424}



\bibitem{fat}  V.A. Fateev, Nucl. Phys. {\bf B479} (1996) 594

\bibitem{fatdevega} V.A. Fateev and H. deVega, \JPA{25}{1992}{2693}, V. Fateev, \IJMPA{6}{1991}{2109}

\bibitem{col}S. Coleman, \PRD{11}{1975}{2088}

\bibitem{man}S. Mandelstam, \PRD{11}{1975}{3026}


\bibitem{mont}C. Montonen and D. Olive, \PLB{72}{1977}{117}

\bibitem{vafa}C. Vafa and E. Witten, \NPB{431}{1994}{3};
N. Seiberg and E. Witten, \NPB{426}{1994}{16}; \NPB{431}{1994}{484}



\bibitem{ferr}L.A. Ferreira, J-L Gervais, J. Sanchez-Guillen and M.V. Saveliev,
\NPB{470}{1996}{236}




 
\bibitem{dubna}J.F. Gomes, E.P. Gueuvoghlanian, G.M. Sotkov and A.H.
Zimerman, ``Classical Integrability of Non Abelian Affine Toda Models, 
to appear in the Proc. of the XXIII
International Colloquium on Group Theoretical Methods in
Physics, Ed. G. Pogosyan et. al., Dubna (2000), hepth/0010257


\end{thebibliography}
\end{document}